\documentclass{emulateapj}

\usepackage{amsmath}
\usepackage{amssymb}
\usepackage{amsfonts}
\usepackage{amstext}
\usepackage{graphicx}
\usepackage{fancybox}
\usepackage{enumitem}
\usepackage[usenames,dvipsnames]{color}
\usepackage{pstricks}
\usepackage{soul, color}
\usepackage{latexsym}
\usepackage{pifont}
\usepackage{shorttoc}
\usepackage{subfigure}
\usepackage{textcomp} 
\usepackage{hyperref}
\usepackage{bm}
\usepackage{lipsum}
\usepackage{threeparttable}
\usepackage{epsfig}

\def\aap{A\&A}
\def\mnras{MNRAS}
\def\apj{ApJ}
\def\apjs{ApJS}
\def\apjl{ApJL}
\def\nat{Nature}

\def\pasp{PASP}
\def\aj{AJ}

\newcommand{\kms}{km s$^{-1}\hspace{1mm}$}

\newcommand{\no}[1]{}%per comentar espais grans

\newcommand{\hs}{\hspace{1mm}}

\def\lsim{~\rlap{$<$}{\lower 1.0ex\hbox{$\sim$}}}

\def\gsim{~\rlap{$>$}{\lower 1.0ex\hbox{$\sim$}}}

\shorttitle{Ly$\alpha$ Signatures from DCBHs}
\shortauthors{Dijkstra et al.}

\begin{document}

\title{Ly$\alpha$ Signatures from Direct Collapse Black Holes}

\author{Mark Dijkstra$^1$, Max Gronke$^1$, and David Sobral$^{2,3}$}
\affil{$^1$Institute of Theoretical Astrophysics, University of Oslo,
P.O. Box 1029 Blindern, N-0315 Oslo, Norway}
\affil{$^2$ Department of Physics, Lancaster University, Lancaster, LA1, 4YB, UK}
\affil{$^3$ Leiden Observatory, Leiden University, P.O. Box 9513, 2300 RA Leiden, The Netherlands}
\altaffiltext{1}{mark.dijkstra@astro.uio.no}

\begin{abstract} 
`Direct collapse black holes' (DCBHs) provide possible seeds for supermassive black holes that exist at redshifts as high as $z\sim 7$. We study Ly$\alpha$ radiative transfer through simplified representations of the DCBH-scenario. We find that gravitational heating of the collapsing cloud gives rise to a Ly$\alpha$ cooling luminosity of up to $\sim 10^{38}(M_{\rm gas}/10^6\hs M_{\odot})^2$ erg s$^{-1}$. The Ly$\alpha$ production rate can be significantly larger during the final stages of collapse, but collisional deexcitation efficiently suppresses the emerging Ly$\alpha$ flux. Photoionization by a central source boosts the Ly$\alpha$ luminosity to $L_{\alpha} \sim 10^{43}(M_{\rm BH}/10^6 M_{\odot})$ erg s$^{-1}$ during specific evolutionary stages of the cloud, where $M_{\rm BH}$ denotes the mass of the black hole powering this source. We predict that the width and velocity off-set of the Ly$\alpha$ spectral line range from a few tens to few thousands km s$^{-1}$, depending sensitively on the evolutionary state of the cloud. 
%The broad Ly$\alpha$ lines would be affected weakly by the intergalactic medium (IGM), which boosts the prospect for their detectability. 
We also compare our predictions to observations of CR7 (Sobral et al. 2015), a luminous Ly$\alpha$ emitter at $z\sim 7$, which is potentially associated with a DCBH. If CR7 is powered by a black hole, then its Ly$\alpha$ flux alone requires that $M_{\rm BH}> 10^7 M_{\odot}$, \textcolor{black}{which exceeds the mass of DCBHs when they first form}. The observed width of the Ly$\alpha$ spectrum favors the presence of only a low column density of hydrogen, $\log [N_{\rm HI}/{\rm cm}^{-2}]\sim 19-20$. The shape of the Ly$\alpha$ spectrum indicates that this gas is outflowing. \textcolor{black}{These requirements imply that if CR7 harbors a DCBH, then the physical conditions that enabled its formation have been mostly erased, which is in agreement with theoretical expectations. }
\end{abstract}

\keywords{cosmology: theory--cosmology: dark ages, reionization, first stars
--quasars: supermassive black holes--galaxies: high-redshift--radiative transfer--scattering}

\section{Introduction}\label{sec:intro}
 The `Direct Collapse Black Holes' (DCBH) presents a channel for forming supermassive black holes (SMBHs) that exist in the Universe at $z\gsim 6$ \citep[see e.g.][for a  review]{JH16}. In the DCBH scenario primordial gas inside dark matter halos with $T_{\rm vir} \geq 10^4$ K collapses {\it directly} into a $\sim 10^4-10^6 M_{\odot}$ black hole, without any intermediate star formation. DCBH formation requires primordial gas to collapse isothermally at $T\sim 10^4$ K \citep[e.g.][]{BL03}, which has been shown to prevent fragmentation \citep[e.g.][]{Li03,Omukai05}. Isothermal collapse at $T\sim 10^4$ K is possible when primordial gas is kept free of molecular hydrogen, H$_2$, for example as a result of a strong photo dissociating radiation background \citep[e.g.][]{BL03,Shang10,WG12,VB13,Sugimura14,Regan14,Fernandez14,Agarwal15,Inayoshi15,Latif15} and/or magnetic fields \citep{Sethi10,VB13}. 

The DCBH formation process is a complex problem which causes theoretical predictions for the number density of DCBHs span to span orders of magnitude \citep[e.g.][]{D14,H16}. For this reason it is extremely valuable to have observational signposts of this process. \citet{Ag13} presented predictions for the broad-band colours of DCBH host galaxies, under the assumption that the spectrum emitted by the accretion disk surrounding the DCBH is a multi-colored disk. Under this assumption, DCBH host galaxies are characterized by blue UV slopes ($\beta \sim -2.3$), which are similar to those predicted for metal poor, young stars. More recently, Dijkstra et al. (2016) have shown that the physical conditions that are required for DCBH formation are optimal for Ly$\alpha$ pumping of the $2p$ level of atomic hydrogen \citep[see][]{Pottasch60,FP61}, and that the $2p$ levels are overpopulated relative to the $2s$ level. These inverted level populations give rise to stimulated $2p_{3/2}\rightarrow 2s_{1/2}$ emission at a rest-frame wavelength of $\lambda=3.04$ cm.  Dijkstra et al. (2016) show that in simplified models of the DCBH scenario, the resulting maser amplifies the background background Cosmic Microwave Background (CMB) by a factor of up to $10^5$, which renders these clouds detectable with the Square Kilometer Array. Dijkstra et al. (2016) performed detailed Ly$\alpha$ radiative transfer calculations to compute their signal. The goal of this paper is to focus on the Ly$\alpha$ line itself, and to demonstrate that there is useful information in both the total Ly$\alpha$ flux and its spectral line shape. The motivation for this work is that 

\begin{itemize}[leftmargin=*]
\item Ly$\alpha$ is theoretically expected to be the most prominent spectral feature in the spectrum emerging from clouds collapsing into or onto a DCBH (see e.g. Inayoshi et al. 2015b, Pallottini et al. 2015, Agarwal et al. 2016). Importantly, the Ly$\alpha$ radiative transfer problem is simplified enormously in the DCBH formation scenario: ordinarily, Ly$\alpha$ transfer through the interstellar medium of galaxies depends sensitively on the distribution, composition and kinematics of cold interstellar gas, all of which are closely connected to stellar feedback processes \citep[see][for extended reviews]{Review,Barnes14,Hayes15}. In contrast, in the DCBH scenario gas fragmentation is suppressed, and no star formation has taken place, which makes stellar feedback (chemical, kinematical) irrelevant. That is, {\it the major challenges that one ordinarily faces when modelling interstellar Ly$\alpha$ transfer are absent in the DCBH-scenario}. It is worth stressing that an immediate consequence of this difference is that the Ly$\alpha$ line emerging from gas in the DCBH scenario is likely different than that emerging from star forming galaxies. 

\textcolor{black}{The physical conditions of the gas that enabled DCBH formation persist briefly after the black hole has formed. Merging with nearby halos occurs over time-scales of tens of Myr \citep[e.g.][also see Tanaka 2014]{Hartwig16}, which would likely mostly erase these conditions. We will also perform Ly$\alpha$ transfer calculations through clouds that are illuminated by a luminous ionizing source, under the assumption that the gas properties are the same as those that enabled DCBH formation. This also provides us with observational signposts of the DCBH formation process.}
\item The recent observation of CR7, a luminous Ly$\alpha$ emitting source at $z\sim 6.6$ (Sobral et al. 2015, Matthee et al. 2015), with a large Ly$\alpha$ equivalent width, and an accompanying luminous, large equivalent width He1640 line \citep{Sobral15}. The fact that HST imaging of CR7 reveals that it is composed of a blue galaxy close to a more massive galaxy has further fueled speculation that CR7 is associated with a DCBH \citep{Sobral15,Pallottini15,AgCR7,Hartwig16}. \textcolor{black}{Previous works have focussed on the observed line fluxes and equivalent widths of Ly$\alpha$ and He1640. The observed ratio of the flux in these lines favor photoionization by sources with very hard spectra (with $T_{\rm eff}>10^5$ K, see Sobral et al. 2015, Hartwig et al. 2016, also see Johnson et al. 2011), and possibly requires a significant fraction of the emitted Ly$\alpha$ flux not to have been detected \citep[][]{Sobral15,AgCR7}.}

Because CR7 has such a bright Ly$\alpha$ emission line, there is excellent data on the Ly$\alpha$ spectral line shape (Sobral et al. 2015). In this paper we demonstrate that it is possible to extract unique physical information from the Ly$\alpha$ line. \textcolor{black}{Moreover, our analysis also highlights the physical ingredients that must be considered when modeling Ly$\alpha$ line emission.}
\end{itemize}

The outline of this paper is as follows. In \S~\ref{sec:model} we describe our model and list the relevant Ly$\alpha$ radiative processes, including Ly$\alpha$ emission processes (\S~\ref{sec:Lya}) and radiative transfer processes (\S~\ref{sec:RT}). We present our main results such as the predicted Ly$\alpha$ luminosity (\S~\ref{sec:lum}) and spectra (\S~\ref{sec:spec}) of clouds collapsing into or onto a DCBH. We compare our predictions to observations of CR7 in \S~\ref{sec:shell_model}, before presenting our conclusions and outlook in \S~\ref{sec:conc}.

\section{Model}
\label{sec:model}
\subsection{Geometry \& Useful Physical Quantities}

Our analysis focusses on spherical and ellipsoidal gas clouds. As we discuss in \S~\ref{sec:lum}, our predicted luminosities are unlikely to be affected by this simplifying assumption. However, our predicted spectra will change (see \S~\ref{sec:spec}). Throughout the paper, we derive numerical values in our equations assuming a uniform gas density profile, which is characterized by a single number density, $n$.  For a given gas mass $M_{\rm gas}$, this gives a cloud radius $R$. Results for uniform clouds are easy to verify and interpret. Adopting a uniform gas density allows us to simplify the Ly$\alpha$ transfer enormously, and occasionally use analytic result from earlier work. Importantly, we show in Appendix~\ref{app:iso} that these results differ only slightly from those from obtained assuming a (cored) isothermal density profile (see e.g. Shang et al. 2010, Pacucci \& Ferrara 2015). In addition, the density profile changes the {\it average} density at which certain processes become important compared to calculations that assume a uniform density, but otherwise does not affect our results. 

For isothermal density profile the gas density becomes very large towards the centre of the collapsing gas cloud, which may lead to the formation of a quasi-star, a supermassive star, or a direct-collapse black hole in the centre of the cloud. We therefore also consider models which contain a central source of ionising radiation, as in Dijkstra et al. (2016). That is, our models assume either that the gas is collapsing {\it into} a black hole, or {\it onto} a central massive black hole that formed out of the inner (denser) gas within the cloud (accretion rates onto the central black hole can be up to $\sim 0.1 M_{\odot}$ yr$^{-1}$, e.g. Latif \& Volonteri 2015). 

\textcolor{black}{As we justified in \S~\ref{sec:intro}, we focus on the emission properties of gas with properties that enabled the DCBH scenario}, i.e. the gas is pristine, no fragmentation occurred, and no stars have formed. For these reasons, we expect our predictions to be most relevant only during the early phase of DCBH formation. As the dark matter halo merger rates increase as $(1+z)^{5/2}$ \citep[e.g.][]{Tanaka14}, the DCBH host halo is expected to merge with another halo on a short time scale \textcolor{black}{\citep[$\sim$ tens of Myr; see][]{Hartwig16}}. In this case, the changes in gas conditions will likely make the Ly$\alpha$ more complex, and more reminiscent of what is happening in ÒordinaryÓ metal-poor star forming galaxies. This will not affect our predicted (intrinsic) Ly$\alpha$ luminosities, but it will affect our predicted spectra (\textcolor{black}{also see \S~\ref{sec:real}}).

Finally, it is useful to recall that the mass of a dark matter halo with virial temperature of $T_{\rm vir}=10^4$ K is $M_{\rm tot}=10^8(\mu/0.6)^{-3/2}([1+z]/11)^{-3/2}$ $M_{\odot}$ ($\mu$ denotes the mean particle mass in units of the proton mass), which has a virial radius of $r_{\rm vir}=1.8([1+z]/11)^{-1}(M_{\rm tot}/10^8 M_{\odot})^{1/3}$ kpc \citep{BL01}. The average number density of hydrogen atoms/nuclei at virialization is $\bar{n}= 0.048([1+z]/11)^3$ cm$^{-3}$.  Linear contraction by a factor of $x$ increases the number density by an additional factor of $x^3$.

\subsection{Ly$\alpha$ Emission}
\label{sec:Lya}

The DCBH scenario involves the collapse of a gas cloud, which cools by via atomic line cooling, $f_{\alpha}\sim 40\%$ of which is in the form of Ly$\alpha$ emission \citep[e.g.][]{Review}. Ly$\alpha$ line cooling compensates for the change in gravitational binding energy, denoted with $U_{\rm bind}$, and is therefore of order
\begin{eqnarray}
& L_{\alpha,{\rm int}}^{\rm g-cool}= f_{\alpha}\frac{dU_{\rm bind}}{dt}=f_{\alpha}\frac{3GM_{\rm gas}^2}{5R^2}\dot{R}  \\ \nonumber
\approx& 4\times 10^{36}\hs{\rm erg\hs s}^{-1} \left( \frac{M_{\rm gas}}{10^6\hs M_{\odot}}\right)^2  \left( \frac{R}{44\hs{\rm pc}}\right)^{-2}  \left( \frac{\dot{R}}{10\hs{\rm km \hs s}^{-1}}\right), \\ \nonumber
\approx& 4\times 10^{36}\hs{\rm erg \hs s}^{-1} \left( \frac{M_{\rm gas}}{10^6\hs M_{\odot}}\right)^2  \left( \frac{n}{100\hs{\rm cm}^{-3}}\right)^{2/3} \left( \frac{\dot{R}}{10\hs{\rm km\hs s}^{-1}}\right) ,
\label{eq:gravheat}
\end{eqnarray} 
where we assumed that dark matter does not contribute to the gravitational potential, which is appropriate when the gas has collapsed to the high densities that we consider in this paper. A gas cloud of mass $M_{\rm gas}=10^{6} M_{\odot}$ with radius $R=44$ pc has a mean enclosed number density of hydrogen atoms that is $\bar{n} \approx 100$ cm$^{-3}$. The total Ly$\alpha$ gravitational cooling luminosity that is {\it produced} increases as $\propto n^{2/3}$. \textcolor{black}{However, as we discuss in \S~\ref{sec:RT} Ly$\alpha$ photons generally are destroyed and do not escape for sufficiently large $n$ (even in the complete absence of dust). The predicted Ly$\alpha$ cooling luminosity that can be observed therefore differs from the produced - also known as the `intrinsic' (hence the subscript `int') - Ly$\alpha$ luminosity.}
 
It is possible to boost the Ly$\alpha$ luminosity if the gas is (kept) ionized by radiation coming from the accretion disk surrounding the central BH (e.g. Haiman \& Rees 2001). The total Ly$\alpha$ luminosity powered by recombinations is 
\begin{eqnarray}
%\begin{aligned}
&L_{\alpha,{\rm int}}^{\rm rec} \approx 0.68 E_{{\rm Ly}\alpha} \alpha_{\rm B}(T)\int dV \hs n_e n_p  \\ \nonumber
&\approx 3\times 10^{41}\hs{\rm erg\hs s}^{-1} \left( \frac{M_{\rm gas}}{10^6\hs M_{\odot}}\right)^2  \left( \frac{R}{44\hs{\rm pc}}\right)^{-3} \left( \frac{T}{10^4\hs {\rm K}}\right)^{-0.7}, \\  \nonumber
&\approx 3\times 10^{41}\hs{\rm erg\hs s}^{-1} \left( \frac{M_{\rm gas}}{10^6\hs M_{\odot}}\right)^2  \left( \frac{n}{100\hs{\rm cm}^{-3}}\right)\left( \frac{T}{10^4\hs {\rm K}}\right)^{-0.7},
%\end{aligned}
\label{eq:rec}
\end{eqnarray} where $\alpha_{\rm B}(T)\approx 2.6\times 10^{-13}(T/10^4\hs{\rm K})^{-0.7}$ cm$^{3}$ s$^{-1}$ denotes the case-B recombination coefficient (in our calculations we take the more accurate approximation from Hui \& Gnedin 1997). The factor `0.68' denotes the number of Ly$\alpha$ photons produced per recombination event (see Dijkstra 2014, for details). The Ly$\alpha$ recombination luminosity is orders of magnitude larger than the signal expected from gravitational heating. The total recombination rate is $\dot{N}_{\rm rec}=L_{\alpha}^{\rm rec} /E_{{\rm Ly}\alpha}$, where $E_{{\rm Ly}\alpha}=10.2$ eV denotes the energy of a Ly$\alpha$ photon. The cloud is only fully ionized when the recombination rate is less than the production rate of ionising photons by the accession disk, $\dot{N}_{\rm ion}$. In the case of $\dot{N}_{\rm rec} > \dot{N}_{\rm ion}$, the HII sphere does not extend to the edge of the cloud, and the ionized gas is `ionization bound'.  In the other case ($\dot{N}_{\rm rec} < \dot{N}_{\rm ion}$), the HII sphere extends right to the edge of the cloud, and the ionized gas is `density bound'.

When the cloud is not fully photoionized, then a significant fraction of the high-energy ionising photons (X-rays) can be absorbed in the neutral gas, which would be then be photoheated. This gas efficiently radiates away this energy, again mostly in Ly$\alpha$. In this case an even larger fraction of the bolometric luminosity of the accretion disk can be converted into Ly$\alpha$ radiation inside the cloud (see \S~\ref{sec:lum} for additional discussion).

\textcolor{black}{When the intense photodissociating radiation field that is required for the DCBH formation scenario is sourced by a nearby star forming galaxy (e.g. Dijkstra et al. 2008), then this galaxy also irradiates the collapsing cloud with X-ray photons \citep{Inayoshi15}. These X-ray photons provide an additional heat source that powers Ly$\alpha$ emission. To estimate this Ly$\alpha$ luminosity we consider a star forming galaxy that provides a photodissociating flux density of $J_{21}=10^3$ ($J_{21}$ denotes the flux density in units of $10^{-21}$ erg s$^{-1}$ cm$^{-2}$ Hz$^{-1}$ sr$^{-1}$). This flux can be provided by a galaxy with SFR$\sim 1 \hs M_{\odot}(d/3\hs {\rm kpc})^2$ yr$^{-1}$, where $d$ is the distance to the galaxy (see Dijkstra et al. 2008). The local relation between SFR and X-ray luminosity (e.g. Mineo et al. 2012, where $L_{\rm X}$ is measured in the 0.5-8.0 keV band) implies that the X-ray luminosity of this galaxy is $L_{\rm X} \sim 3\times 10^{39}(d/3\hs {\rm kpc})^2$ erg s$^{-1}$. The total luminosity incident on the cloud of radius $R$ equals $L_{\rm X}\sim 3\times 10^{36}(R/100\hs {\rm pc})^{-2}$ erg s$^{-1}$, which can become comparable to the gravitational cooling luminosity during early stages of the collapse.}

\subsection{Ly$\alpha$ Transfer}
\label{sec:RT}

Gas with HI column densities in excess of $10^{20}$ cm$^{-2}$ is considered `extremely' optically thick to Ly$\alpha$ radiation \citep[e.g.][]{Adams72,Neufeld90}. Ly$\alpha$ transfer through extremely opaque media has been studied for decades \citep[see][for an extended review]{Review}. The line-center optical depth, $\tau_0$, of gas with an HI column density $N_{\rm HI}$ is:
\begin{equation}
\tau_0=5.9\times 10^6 \left(\frac{N_{\rm HI}}{10^{20}\hs{\rm cm}^{-2}} \right) \left( \frac{T}{10^4\hs{\rm K}}\right)^{-1/2},
\label{eq:tau0}
\end{equation} where $T$ denotes the gas temperature. Ly$\alpha$ photons typically scatter\footnote{\textcolor{black}{The {\it trapping time} of Ly$\alpha$ photons equals $t_{\rm trap}=Bt_{\rm cross}$, where $t_{\rm cross}\sim R/c$ denotes the light crossing time, and $B\approx 12(N_{\rm HI}/10^{20}\hs{\rm cm^{-2}})^{1/3}(T/10^4\hs{\rm K})^{-1/3}$ \citep{Adams75,D16}. For typical cloud sizes of $\sim 1-100$ pc, and $\log \tau_0\sim 8-12$, we have $t_{\rm trap}\sim 10^3-10^4$ yrs (longer trapping times for lower densities), which is significantly less than the collapse time of the halo. While Ly$\alpha$ photons undergo a tremendous number of scattering events, their escape is not affected by them getting `trapped' in the scattering medium. This is because the vast majority of scattering events occurs in the line core, where the mean time is very short, $t_{\rm mfp}=(cn_{\rm HI}\sigma_0)^{-1} \sim 1(n_{\rm HI}/10^3\hs{\rm cm}^{-3})^{-1}$ s.} } of order $N_{\rm scat} \sim \tau_0$ before escaping from a static medium \citep{Adams72,Harrington73,Neufeld90}. 

The number of scattering events can be reduced tremendously in multiphase media and/or media in which velocity gradients are present. In the DCBH formation scenario, fragmentation is suppressed and we expect the gas distribution to be smooth. Velocity gradients only reduce $N_{\rm scat}$ by a factor of order $\sim \Delta v/v_{\rm th}$, where $v_{\rm th}=12.9(T/10^4\hs{\rm K})^{1/2}$ km s$^{-1}$ is the thermal velocity of hydrogen and $\Delta v$ denotes the (maximum) velocity difference between the centre and the edge of the cloud (Bonilha et al. 1979, also see Dijkstra et al. 2016 for a more quantitative analysis).

Ly$\alpha$ scattering is limited by a number of other processes, which we briefly describe here (for a more detailed description the reader is referred to Dijkstra et al. 2016). Ly$\alpha$ photons can be destroyed in the following ways:
\begin{enumerate}[leftmargin=*] 
\item Molecular hydrogen (H$_2$) has two transitions that lie close to the Ly$\alpha$ resonance: ({\it a}) the $v=1-2P(5)$ transition, which lies $\Delta v=99$ km s$^{-1}$ redward of the Ly$\alpha$ resonance, and ({\it b}) the $1-2R(6)$ transition which lies $\Delta v=15$ km s$^{-1}$ redward of the Ly$\alpha$ resonance. Vibrationally excited $H_2$ may therefore convert Ly$\alpha$ photons into photons in the $H_2$ Lyman bands  \citep[][and references therein]{Neufeld90}, and thus effectively destroy Ly$\alpha$.
\item Photoionization by Ly$\alpha$ of the (tiny) fraction of atoms in the $2p$ and $2s$ states.
\item Ly$\alpha$ photons can detach the electron from the H$^-$ ion. This process is (slightly) less important than destruction via photoionization.
\item Induced transitions $2p\rightarrow 2s$ by the CMB can be important when the $2s$ and $2p$ level populations are inverted. These induced transitions remove Ly$\alpha$ photons (see Dijkstra et al. 2016). This effect however, can become important at higher densities where $f_{\rm esc}$ is already tiny due to collisions (see the next process).
\item Collisions between a proton and the atom in the excited $2p$ state which put it in the $2s$ state, at which point radiative transitions to the ground state are only permitted via emission of two continuum photons (whose combined energy is 10.2 eV). This last process is most important in limiting the escape fraction of Ly$\alpha$ photons (see Dijkstra et al. 2016), and we ignore the other processes hereafter.
\end{enumerate}

The lifetime of the hydrogen atom in the $2p$-state is $t_{2p}=A_{\alpha}^{-1}$, where $A_{\alpha}=6.25\times 10^8$ s$^{-1}$ denotes the Einstein-A coefficient of the Ly$\alpha$ transition. The collisional deexcitation rate (in s$^{-1}$) is $n_pq_{2p2s}$, where $n_p$ denotes the number density of free protons and $q_{2p2s}=1.8\times 10^{-4}$ cm$^3$ s$^{-1}$ denotes the $2p \rightarrow 2s$ `collision strength', which depends only weakly on temperature \citep[e.g.][]{D05}. The probability that a Ly$\alpha$ photon is eliminated via a collisional deexcitation event is then 
\begin{equation}
P_{\rm dest}(n,T)=\frac{n_p(T)q_{2p2s}(T)}{n_p(T)q_{2p2s}(T)+A_{\alpha}}. 
\label{eq:pdest}
\end{equation}
The fraction of Ly$\alpha$ photons that escape, $f_{\rm esc}$, equals the probability that a Ly$\alpha$ photon is {\it not} destroyed after $N_{\rm scat}$ scattering events. The Ly$\alpha$ escape fraction therefore equals
\begin{equation}
f_{\rm esc}(n,T)=\left[1-P_{\rm dest}(n,T)\right]^{N_{\rm scat}}.
\label{eq:fesc}
\end{equation} Because $P_{\rm dest} \ll 1$, we can approximate Eq~\ref{eq:fesc} as $f_{\rm esc}\approx 1-N_{\rm scat}P_{\rm dest}$. We can obtain the maximum number of scattering events, $N^{\rm max}_{\rm scat}$, by setting $f_{\rm esc}=0$, which yields
\begin{equation}
N^{\rm max}_{\rm scat}=\frac{1}{P_{\rm dest}}=\frac{n_pq_{2p2s}+A_{\alpha}}{n_pq_{2p2s}}\underset{T=10^4{\rm K}}{\approx} 10^{13}\left( \frac{n}{50\hs{\rm cm}^{-3}}\right)^{-1}.
\end{equation} Here, we assumed collisional ionization equilibrium. Under this assumption, $n_p=3.5\times10^{-3}n$ for $T=10^4$ K. These numbers depend strongly on temperature: for $T=8000$ K [$T=6000$ K], $N^{\rm max}_{\rm scat}$ is increased by two [five] orders of magnitude. The temperature dependence of $N^{\rm max}_{\rm scat}$ is shown in Figure~\ref{fig:uniformcloud} in the Appendix. We take this temperature dependence into account when we present our results in \S~\ref{sec:lum}, by matching the cooling rate of the gas (which depends strongly on $T$) to the heating rate. 

It is worth stressing that $N^{\rm max}_{\rm scat}$ {\it decreases} as the cloud contracts as $N^{\rm max}_{\rm scat} \propto n^{-1}_{\rm p} \propto n^{-0.9}$ (as $n_{\rm p} \propto n^{y}$ with $y=0.88$ to a good approximation, see Appendix~\ref{app:intermediate}). For comparison, the cloud column density - and hence the line centre optical depth, and consequently the average number of scatterings before escape - {\it increases} as $N_{\rm HI} \propto n^{2/3}$. This implies that there exists a critical density of the cloud, $n_{\rm max}$, above which it abruptly stops emitting Ly$\alpha$ radiation. 
In \S~\ref{sec:lum} for example we find that collisional de-excitation becomes important at $n_{\rm max}\sim 10^{4}-10^{6}$ cm$^{-3}$, where the precise number depends on the gas temperature.
\begin{figure*}
\vspace{-5mm}
\centering
\includegraphics[width=10.0cm,angle=270]{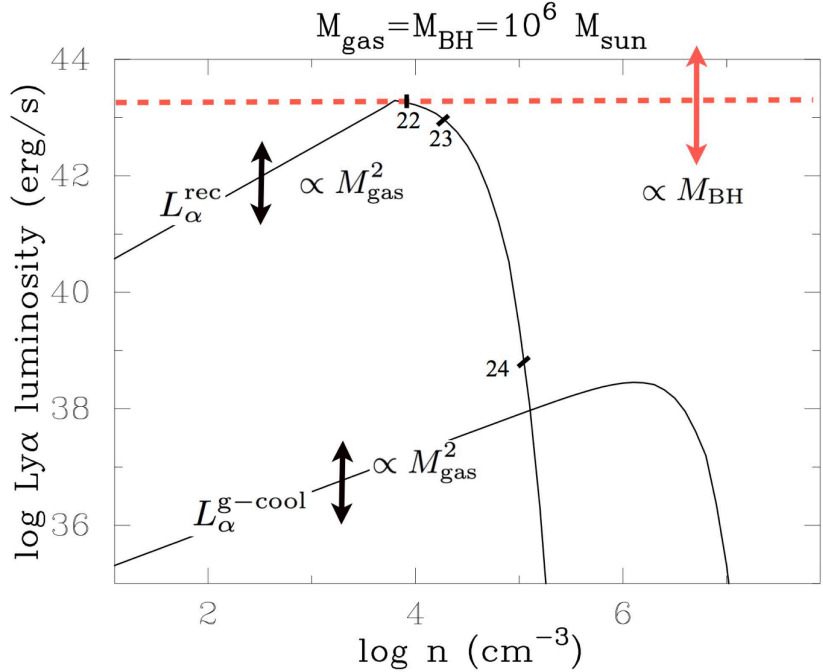}
\vspace{0mm}
\caption[]{This Figure shows the total Ly$\alpha$ luminosity emerging from a (pristine) gas cloud of mass $M_{\rm gas}=10^{6}M_{\odot}$ collapsing into a black hole, as a function of its (average) number density of hydrogen. The lower line shows $L_{\alpha}^{\rm g-cool}$, the Ly$\alpha$ cooling luminosity in response to gravitational heating. This cooling luminosity peaks at $L_{\alpha}^{\rm g-cool}\sim 10^{38}(M_{\rm gas}/10^6\hs M_{\odot})^2$ erg s$^{-1}$ at a number density of $n\sim 10^6$ cm$^{-3}$. For higher densities, the HI column density grows so large that collisional deexcitation of hydrogen atoms in the $2p$ suppresses the emerging Ly$\alpha$ flux (see Eq~\ref{eq:pdest} and Eq~\ref{eq:fesc}). The Ly$\alpha$ luminosity is boosted when the gas is irradiated by a central source powered by an already formed DCBH of mass $M_{\rm BH}$. The Ly$\alpha$ recombination scales as $\propto M^2_{\rm gas}$ up to a maximum $L^{\rm rec}_{\alpha} \sim 10^{43}(M_{\rm BH}/10^6\hs M_{\odot})$ erg s$^{-1}$. This maximum ({\it red dashed horizontal line}) is reached over a limited range of densities and is independent of $M_{\rm gas}$. For lower densities the ionized gas is density-bound, which reduces the conversion efficiency of ionising radiation into Ly$\alpha$. For higher densities the cloud is ionization-bound, and collisional deexcitation again reduces the Ly$\alpha$ luminosity. Collisional deexcitation kicks in at lower density than in the pure gravitational heating case (see text).  Labels on the $L_{\alpha}^{\rm rec}$-curve indicate the log of the HI column density and labeled arrows mark proportionalities.}
\label{fig:lum}
\end{figure*} 

Ly$\alpha$ radiative transfer through extremely optically thick gas can be described accurately as a diffusion process in both real and frequency space \citep[e.g.][]{R94,HM12}. That is, as the photons scatter diffuse outwards through the cloud, they also diffuse away from line centre, which facilitates their escape. Scattering thus broadens the spectral distribution of Ly$\alpha$ photons. The characteristic broadening of the line through a static medium is \citep[][]{Adams72,Harrington73,Neufeld90}
\begin{eqnarray}
\frac{{\rm FWHM }}{10^3\hs{\rm km\hs s}^{-1}}\approx 1.5\Big{(}\frac{N_{\rm HI}}{10^{22}\hs{\rm cm^{-2}}} \Big{)}^{1/3}\Big{(} \frac{T}{10^4\hs{\rm K}}\Big{)}^{1/6} \\ \nonumber
\approx 1.5\left( \frac{n}{50\hs{\rm cm}^{-3}}\right)^{2/9}\left( \frac{M_{\rm gas}}{10^6M_{\odot}}\right)^{1/9} \left(\frac{T}{10^4\hs{\rm K}}\right)^{1/6}.
\label{eq:FWHM}
\end{eqnarray}
For static gas clouds the line is centered on and symmetric around the Ly$\alpha$ resonance. For contracting gas clouds however, the Ly$\alpha$ line is blue shifted relative to the line centre by an amount that depends on the HI column density and the infall velocity profile \citep[e.g.][]{Zheng02,D06}. In \S~\ref{sec:spec} we will present predictions for the line profile, and compare these to observations of CR7 in \S~\ref{sec:shell_model}.\\

\section{Results I: Ly$\alpha$ Luminosity}
\label{sec:lum}
Figure~\ref{fig:lum} shows the Ly$\alpha$ luminosity as a function of density $n$. This plot contains two lines:

$\bullet$ {\bf Gravitational Cooling Luminosity.} For a uniform gas density profile, the gravitational binding energy $U_{\rm bind}=-\frac{3GM^2}{5R}$. At each $n$ we compute the gas temperature $T$ by setting the gravitational heating rate equal to the total cooling rate, i.e. $\Big|\frac{dU_{\rm bind}}{dt}\Big|=\int dV n_e (T) n_{\rm HI}(T)C(T)$, in which $C(T)$ denotes the total gas cooling rate (which includes the relevant cooling rates taken from Hui \& Gnedin 1997). Once we have determined $T$, we compute $P_{\rm dest}(n,T)$, and $f_{\rm esc}(n,T)$ (see Eq~\ref{eq:pdest} and Eq~\ref{eq:fesc}). Figure~\ref{fig:lum} shows the total Ly$\alpha$ cooling luminosity 
\begin{equation}
L_{\alpha}^{\rm g-cool}=f_{\rm esc}(T,n)L_{\alpha,{\rm int}}^{\rm g-cool}.
\end{equation}
Figure~\ref{fig:lum} shows that the gravitational cooling signal reaches a maximum\footnote{The {\it maximum} Ly$\alpha$ cooling luminosity scales as $\propto M_{\rm gas}^{1.86}$, because $N_{\rm scat} \approx \tau_0 \propto N_{\rm HI}=nR \propto n^{2/3}M_{\rm gas}^{1/3}$, while $N^{\rm max}_{\rm scat}\propto n_{\rm p}^{-1} \propto n^{-y}$, where $y \approx 0.88$ (see Appendix~\ref{app:intermediate}). If we set $N_{\rm scat}=N^{\rm max}_{\rm scat}$ then it follows that $n_{\rm max} \propto M_{\rm gas}^{-[3y+2]^{-1}}\propto M_{\rm gas}^{-0.21}$, and the {\it maximum} Ly$\alpha$ cooling $L_{\alpha}^{\rm g-cool} \propto M^2_{\rm gas}n^{2/3}_{\rm max} \propto M_{\rm gas}^{1.85}$.} of $L_{\alpha}^{\rm g-cool}\sim 10^{38}$ erg s$^{-1}$ at a critical density (which we introduced in \S~\ref{sec:RT}) $n_{\rm max}\approx 10^{6}$ cm$^{-3}$, beyond which it rapidly decreases. This rapid decrease is due to the rapid decrease in $f_{\rm esc}(T,n)$.
Intermediate results of our calculations are shown in Figure~\ref{fig:uniformcloud} in Appendix~\ref{app:intermediate}. 

$\bullet$ {\bf Recombination Radiation.} \textcolor{black}{The predicted Ly$\alpha$ luminosity powered by recombination events equals:
\begin{equation}
L_{\alpha}^{\rm rec}=f_{\rm esc}(T,n)\times \left\{ \begin{array}{ll}
         \ L_{\alpha,{\rm int}}^{\rm rec},& L_{\alpha,{\rm int}}^{\rm rec} \leq L_{\alpha,{\rm max}}^{\rm rec}\\
         \ L_{\alpha,{\rm max}}^{\rm rec},& L_{\alpha,{\rm int}}^{\rm rec} > L_{\alpha,{\rm max}}^{\rm rec},\end{array} 
\right..
\end{equation} where the intrinsic Ly$\alpha$ recombination rate $L_{\alpha,{\rm int}}^{\rm rec}$ is given in Eq~2. The Ly$\alpha$ recombination luminosity is bound by the total photoionization rate in the cloud, and hence by the total ionising photon production rate around the black hole. We assumed that the accretion disk around the DCBH is radiating at Eddington luminosity, and that its spectrum is identical to that of unobscured, radio-quiet quasars. Under these assumptions, the total ionising photon emission rate is \citep{Bolton11}
\begin{equation}
\dot{N}_{\rm ion}=6.5\times 10^{53} \left(\frac{M_{\rm BH}}{10^6\hs M_{\odot}} \right)\hs{\rm s}^{-1}.
\end{equation}} The Ly$\alpha$ recombination luminosity then saturates at
\begin{eqnarray}\label{eq:reclya}
L_{\alpha,{\rm max}}^{\rm rec}=F_{{\rm Ly}\alpha} E_{{\rm Ly}\alpha} \dot{N}_{\rm ion}\sim \\ \nonumber
 2 \times10^{43}\left(\frac{F_{{\rm Ly}\alpha}}{2.0}\right)\left(\frac{M_{\rm BH}}{10^6\hs M_{\odot}} \right)\hs{\rm erg} \hs{\rm s}^{-1}.
\end{eqnarray} Here, $F_{{\rm Ly}\alpha}$ denotes the mean number of Ly$\alpha$ photons that are produced per ionizing photon. If each ionizing photon ionized one hydrogen atom, then $F_{{\rm Ly}\alpha}\approx 0.68$ (as we mentioned above). However, photoionization by high-energy photons creates energetic electrons which can collisionally ionize additional atoms\footnote{For example, a 1 keV photon can ionize 25 hydrogen atoms in a neutral gas.}. The boost in the production rate of Ly$\alpha$ photons scales with the mean photon energy above 13.6 eV (Raiter et al. 2010). Our fiducial model assumes a boost by a factor of $\sim 3$, and that therefore $F_{{\rm Ly}\alpha}\sim 3 \times 0.68 \sim 2$. 

Figure~\ref{fig:lum} shows $L_{\alpha}^{\rm rec}$ as a function of density $n$ . This maximum luminosity\footnote{Technically, the assumption $F_{{\rm Ly}\alpha}=2$ assumes that higher energy photons contribute to Ly$\alpha$ production. However, large HI column densities are required to absorb these photons. These higher energy photons are thus only absorbed only after the cloud has become ionization bound, i.e. when the ionized gas is surrounded by neutral gas. The factor $F_{{\rm Ly}\alpha}$ thus formally increases from $F_{{\rm Ly}\alpha}=0.68$ to larger values only after the cloud becomes ionization bound.} $L_{\alpha,{\rm max}}^{\rm rec}$ is reached at a density of $n \sim 10^4$ cm$^{-3}$. Figure~\ref{fig:lum} shows that the recombination luminosity plummets for $n \gsim 10^4$ cm$^{-3}$, which is due the reduction in $f_{\rm esc}(n,T)$ as a result of \textcolor{black}{X-Ray heating: The Ly$\alpha$ escape fraction goes to zero at lower density compared to the case of gravitational heating, because higher energy photons can penetrate the neutral gas which increases its temperature. We compute this temperature by setting the total cooling rate of the gas equal to the total radiative heating rate, which we obtain by assuming that a fraction $f_{\rm heat}$ of the X-ray luminosity goes into heating.} The total \textcolor{black}{X-ray} heating rate is then
\begin{eqnarray}
H^{\gamma}& = & f_{\rm heat}f_{\rm X}L_{\rm edd} \\ \nonumber
&=&1.3\times 10^{43}\left( \frac{f_{\rm heat}}{1.0}\right)\left(\frac{f_{\rm X}}{0.1}\right)\left( \frac{M_{\rm BH}}{10^6 \hs M_{\odot}}\right)\hs {\rm erg}\hs{\rm s}^{-1}, 
\label{eq:gheat}
\end{eqnarray} where the choice $f_{\rm X}\sim 10\%$ is based on observations of luminous quasars for which 10\% of  their total bolometric luminosity is in the 0.5-10 keV band \citep[e.g.][]{Marconi04,Lusso12}. The choice $f_{\rm heat}=100\%$ is high. The enhanced temperature of this photo heated neutral gas gives rise to a larger residual ionized fraction, which enhances collisional deexcitation. We have verified that the choice $f_{\rm heat}=10\%$ would result in a lower equilibrium temperature, a lower residual neutral fraction, which in turn would cause $f_{\rm esc}$ to drop to zero at a higher density, but otherwise leave our results unchanged. 

The previous discussion and Figure~\ref{fig:lum} highlights that we can only observe Ly$\alpha$ signatures from DCBH in cases in which accretion onto the DCBH is already occurring, as the Ly$\alpha$ cooling signal in response to gravitational heating is orders of magnitude fainter than what can be detected now, and in the near future (see e.g. Dijkstra 2014). To reach detectable Ly$\alpha$ luminosities in excess of $L \sim 10^{42}$ erg s$^{-1}$ we need the gas surrounding the central massive DCBH ($M\gsim 10^5 M_{\odot}$ ) to be within a very specific density range. In the next section, we predict the spectrum of Ly$\alpha$ photons emerging from gas in this particular state.

\textcolor{black}{It is worth stressing that our predicted luminosities are set entirely by the energy considerations: the total cooling rate is set by the rate of gravitational heating. The total binding energy of the gas changes for more realistic/complex gas distributions, but only by a factor of order unity. The total recombination rate of Ly$\alpha$ power is limited by the rate at which ionizing photons are emitted into the gas, irrespective of the gas distribution. The gas distribution affects the mean density at which the escape fractions go to zero for both processes, and at which the {\it intrinsic} Ly$\alpha$ recombination rate saturates. The gas distribution more strongly affects the emerging spectra, as we discuss in \S~\ref{sec:spec}.}

\section{Results II: Ly$\alpha$ Spectra}
\label{sec:spec}

Predicting the emerging Ly$\alpha$ spectrum from the clouds is more difficult than predicting the Ly$\alpha$ luminosity, as they - unlike our predictions for luminosity - depend more strongly on the adopted geometry. In this section we focus on the spectra emerging from different sets of models. We show calculations of Ly$\alpha$ spectra emerging from a suite of spherically symmetric, collapsing gas clouds in \S~\ref{sec:cloudnoigm}. We show how these spectra are affected by subsequent radiative transfer in the intergalactic medium in \S~\ref{sec:igm}. We study the directionally dependent spectra emerging from flattened clouds without IGM in \S~\ref{sec:geom_orient} and with IGM in \S~\ref{sec:geom_orient_IGM}. \textcolor{black}{We briefly discuss how more complex gas distributions affect our results in \S~\ref{sec:real}.}

\subsection{Spherical Clouds}\label{sec:cloudnoigm}
\begin{figure}
\includegraphics[width=6.5cm,angle=270]{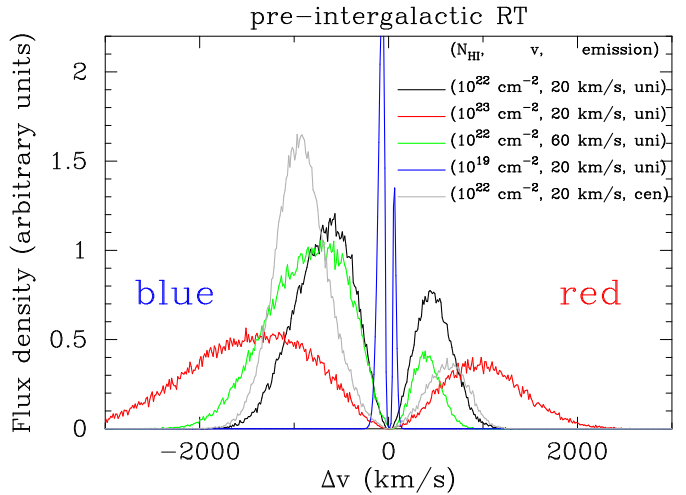}
\caption{Spectra of Ly$\alpha$ photons emerging from gas clouds collapsing into/onto a DCBH. The clouds are characterized by their HI column density ($N_{\rm HI}$), contraction velocity ($v$), and emission profile (uniform or central). These plots show that a generic prediction of the models is the Ly$\alpha$ line is double peaked, with an enhanced blue peak compared to the red. Quantative predictions on the velocity off-set of the peaks depends on $N_{\rm HI}$, $v$ and emissivity profile. The IGM will significantly suppress the blue peak of the line (see Fig~\ref{fig:spec2}).} 
\label{fig:spec1}
\end{figure}

Figure~\ref{fig:spec1} shows Ly$\alpha$ spectra emerging from spherical clouds characterized by ({\it i}) the HI column density ($N_{\rm HI}$), ({\it ii}) a `Hubble-like' contraction of the form $v \propto R$, with a maximum contraction velocity $v_{\rm coll}$ at the edge of the cloud, ({\it iii}) the emissivity profile. We assume either a uniform Ly$\alpha$ emissivity, or a central source fully surrounded by the gas. In all models we set $T=10^4$ K, and that all photons are emitted at line centre. These assumptions do not affect our calculations at all. We discuss each model below:
\begin{figure*}
\centering
\includegraphics[width=8.0cm,angle=270]{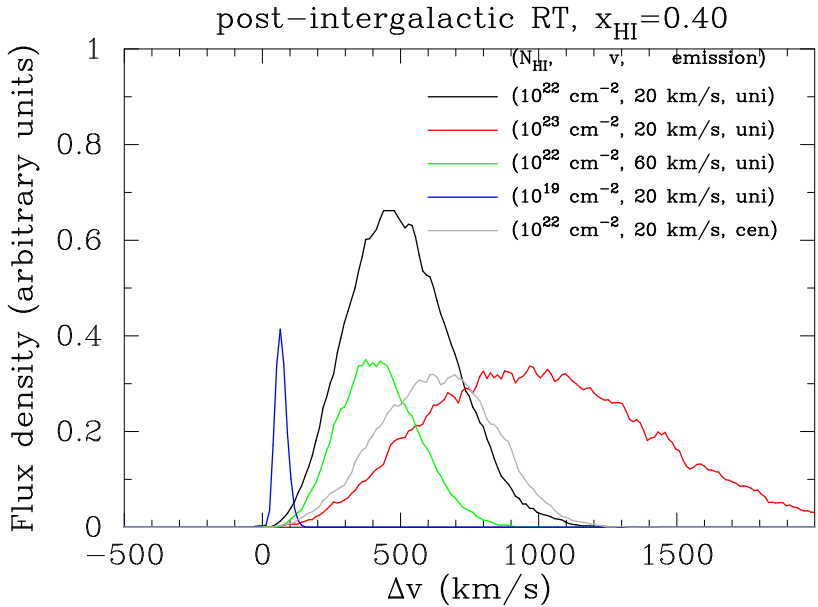}
\caption{Predicted Ly$\alpha$ spectra from Figure~\ref{fig:spec1} after processing these lines through the IGM, using the IGM transmission models from \citet{D11} which accounts for an inhomogeneous distribution of diffuse, neutral intergalactic gas that is associated with patchy reionization. These calculations assumed a volume filling factor of neutral gas of $\langle x_{\rm HI} \rangle =0.40$. This figure shows that the IGM removes the more prominent blue peak of the Ly$\alpha$ spectrum, and leaves the red peak. Processing through the IGM results in redshifted asymmetric Ly$\alpha$ lines, which look remarkably like those produced by scattering through galactic outflows, though predicted redshifts and FWHMs of order $\sim 1000$ km s$^{-1}$ are larger than what has been observed for ordinary Ly$\alpha$ emitting galaxies.} 
\label{fig:spec2}
\end{figure*}
\begin{itemize}[leftmargin=*]

\item The {\it black solid line} shows a model with $N_{\rm HI}=10^{22}$ cm$^{-2}$, $v_{\rm coll}=20$ km s$^{-1}$, and uniform emission. The column density is on the lower end of values associated with the maximum Ly$\alpha$ luminosity in Figure~\ref{fig:lum}. We have chosen $v_{\rm coll}=20$ km s$^{-1}$ (instead of $v_{\rm coll}=10$ km s$^{-1}$ that was adopted in Eq~1) to better illustrate the impact of collapse on the emerging line profile. This spectrum is broadened into the characteristic double peaked profile that is associated with scattering through static, extremely opaque media. Infall enhances the blue-peak compared to the red as expected (see \S~\ref{sec:RT}).

\item The {\it red line} shows a model with $N_{\rm HI}=10^{23}$ cm$^{-2}$, $v_{\rm coll}=20$ km s$^{-1}$, and uniform emission. The boost in HI column density by a factor of ten increases the line centre optical depth by the same factor, which further broadens the line by a factor $10^{1/3} \sim 2.2$ (see Eq~\ref{eq:FWHM}), which is consistent with the broadening shown in Figure~\ref{fig:spec1}. 

\item The {\it blue line} shows a model with $N_{\rm HI}=10^{19}$ cm$^{-2}$, $v_{\rm coll}=20$ km s$^{-1}$, and uniform emission. This model represents the special moment in which the ionized region is ionization bound, but only by a tiny shell of neutral gas. In this case, scattering still gives rise to the double peaked profile, but the peaks are separated only by $\Delta v \sim 130$ km s$^{-1}$. This spectrum can also be thought of as representing the spectrum when the cloud is seen along a low-column density sightline. We return to discuss this in \S~\ref{sec:geom_orient}.

\item The {\it green line} shows a model with $N_{\rm HI}=10^{22}$ cm$^{-2}$, $v_{\rm coll}=60$ km s$^{-1}$, and uniform emission. The main purpose of this plot was to show that the contraction velocity has no major impact on our results. The enhanced contraction velocity enhances the blue peak relative to the red peak. The overall shape of the peaks is mostly preserved.

\item The {\it grey line} shows a model with $N_{\rm HI}=10^{22}$ cm$^{-2}$, $v_{\rm coll}=20$ km s$^{-1}$, and central emission. This plot shows that the precise radial emissivity profile also affects the relative importance of the blue and red peaks, in a similar way as the contraction velocity.
\end{itemize}

These plots show that a generic prediction of the models is a double peaked line profile, with an enhanced blue peak. The precise FWHM of the line depends most strongly on $N_{\rm HI}$ (and hence, the evolutionary stage of the cloud) closely in line with expectations from analytic arguments, but also on the Ly$\alpha$ emissivity profile, and infall velocity profile. At face value, these predicted spectra differ greatly from line profiles that are commonly observed in Ly$\alpha$ emitting galaxies (including CR7, see discussion in \S~\ref{sec:igm}), which typically show asymmetric, redshifted line profiles, indicative of scattering through outflowing gas (see Dijkstra 2014, Hayes 2015 for reviews). However, we stress that the intergalactic medium (IGM) strongly affects especially those photons that emerge on the blue side of the line. We model the impact of the IGM below.

\subsection{Spherical Clouds, with IGM}
\label{sec:igm}
The IGM is optically thick to photons emerging blue ward of the Ly$\alpha$ resonance at\footnote{More precisely, the `effective' optical depth in the Ly$\alpha$ forest exceeds unity, $\tau_{\rm eff} > 1$ at $z \gsim 4$ \citep[see e.g Fig~3 of][]{FG08}} $z \gsim 4$. To first order, the IGM transmits all flux redward of the Ly$\alpha$ resonance, and suppresses the flux blueward of it. In reality, intergalactic radiative transfer of Ly$\alpha$ photons is more complicated as galaxies reside in overdense regions of the Universe, which is more opaque to Ly$\alpha$ radiation than what has been inferred from the Ly$\alpha$ forest observations \citep[e.g.][]{IGM,Laursen11}. In addition, the opacity of the IGM is highest at frequencies close to the Ly$\alpha$ resonance (see Laursen et al. 2011), which makes spectrally narrower emission lines centered on line centre more subject to intergalactic radiative transfer (also see Zheng et al. 2010).
\begin{figure*}
\centering
\includegraphics[width=12.0truecm,angle=270]{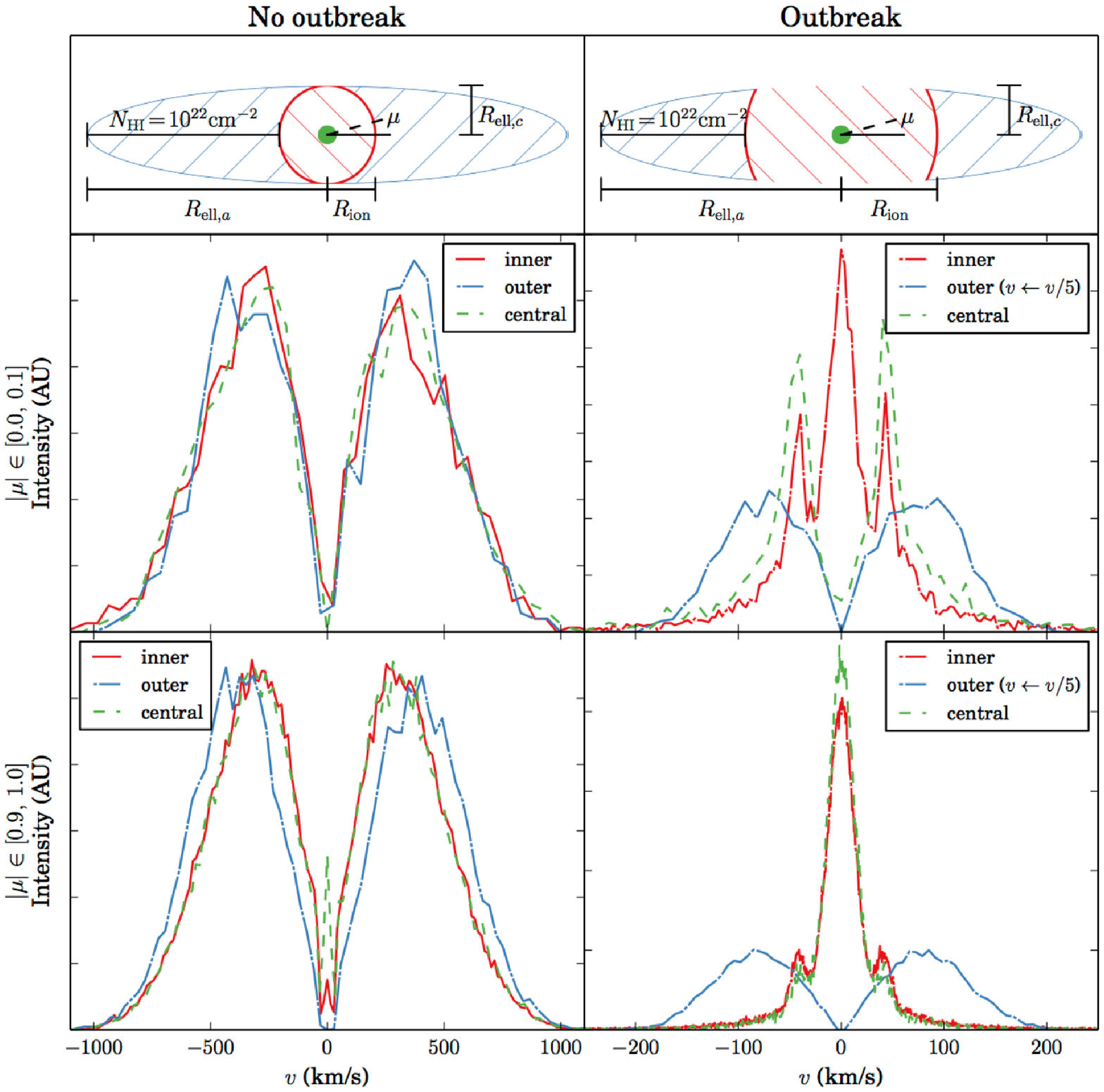}
\caption{The Ly$\alpha$ spectrum arising from an ionized sphere embedded in a ellipsoid filled with hydrogen. The \textit{left column} shows spectra from the `no break-out' (i.e., $R_{\rm ion}= R_{{\rm ell},c}$) scenario whereas the \textit{right column} shows spectra from the `break-out' setup ($R_{\rm ion} = 2 R_{{\rm ell},c}$) as sketched in the panels placed in the \textit{top row}. The observer is placed face-on (so that the column density to the center is maximized) in the \textit{central row} and edge-on (minimizing the column density) in the \textit{bottom row}. In each panel we show the spectra resulting from three different sources indicated in the sketches with the corresponding color: a central source (green dashed line), a uniform luminosity within the ionized region (red solid line), and, a uniform luminosity in the region where the HI is placed (blue broken line). Note, that this curve is rescaled in the `break-out' scenario by a factor of $5$ due to representation purposes. See \S\ref{sec:geom_orient} for details.}
\label{fig:ellipsoid}
\end{figure*}
We approximate intergalactic RT at the redshifts of DCBH formation ($z > 6$) using the models of \citet{D11} which suppress all flux blue ward, {\it and} up to $\sim 100$ km s$^{-1}$ redward of the Ly$\alpha$ resonance. This latter accounts for infall of intergalactic gas onto the dark matter halo that hosts the cloud \citep{IGM,Iliev08}. The model of \citet{D11} accounts for the inhomogeneous distribution of diffuse neutral intergalactic gas that is characteristic of a patchy reionization process\footnote{The models of \citet{D11} account for the inhomogeneous nature of the reionization process. Their IGM transmission curves in these models were calculated for massive, $M\sim 10^{10}-10^{11} M_{\odot}$, dark matter halos. While DCBH occurs in much less massive halos, the requirement of having a luminous nearby neighbour typically places these halos in close proximity to a more massive halo \citep[][but see Agarwal et al. 2012, Visbal et al. 2014 where the LW flux is provided by low-mass nearby galaxies]{D08}.}. For illustration purposes, we assumed that the volume filling factor of the diffuse neutral IGM is $\langle x_{\rm HI} \rangle_V =40\%$, a value that is preferred at $z\sim 7$ by the observed reduction of Ly$\alpha$ flux from galaxies at $z>6$ \citep[see e.g.][Mesinger et al. 2015, Choudhury et al. 2015, \textcolor{black}{this includes both drop-out galaxies and Ly$\alpha$ emitters, see Dijkstra et al. 2014b}]{Review}.

Figure~\ref{fig:spec2} shows the spectra after we have processed the predicted lines through the IGM. The prominent blue peaks are completely eliminated in all cases, which leaves only the red peaks. It is remarkable that processing through the IGM thus results in redshifted asymmetric Ly$\alpha$ lines, which resemble line profiles associated with scattering through galactic outflows. A difference is that the predicted velocity off-set and FWHM of the line can be of order $\sim 1000$ km s$^{-1}$, which is larger than what is typically found for high-redshift Ly$\alpha$ emitting galaxies (see Fig~9 of Willott et al. 2015). \textcolor{black}{It may make it challenging observationally to detect the broadest lines, as (when observed from the ground) they would possibly be spread over multiple OH lines.} It is worth stressing that in all these models the Ly$\alpha$ photons that scattered in the IGM form diffuse, extended Ly$\alpha$ halos surrounding the cloud. \textcolor{black}{The halos would extend significantly further than what has been detected so far.} 

\begin{figure*}
\centering
\vspace{-5mm}
\includegraphics[width=12.0truecm,angle=270]{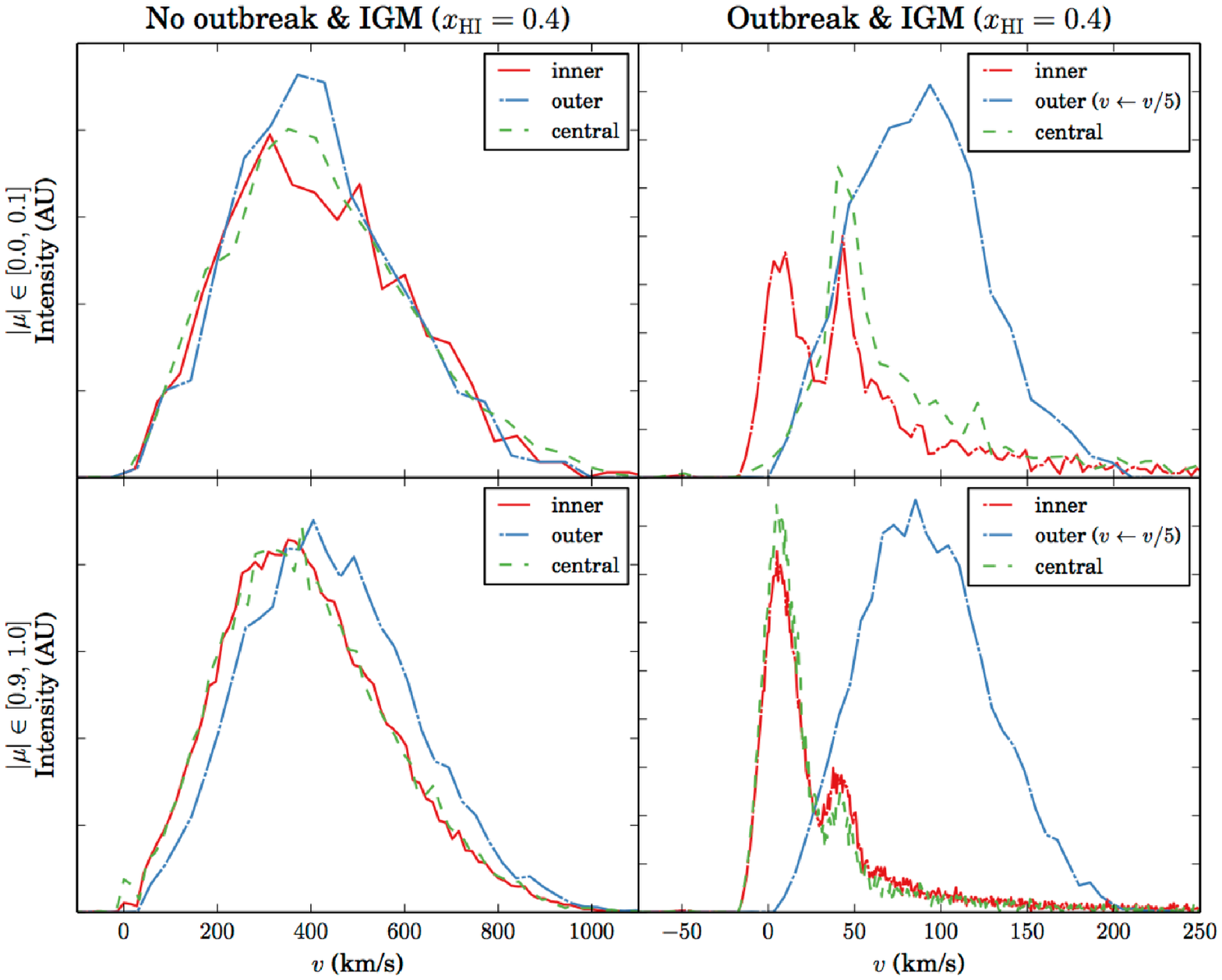}
\vspace{-15mm}
\caption{Same as Figure~\ref{fig:ellipsoid}, but after processing the lines through the IGM (as in Fig~\ref{fig:spec2}).}
\label{fig:ellipsoidigm}
\end{figure*}

\subsection{Ellipsoidal Clouds}
\label{sec:geom_orient}

Our previous calculations assumed spherical symmetry. Here, we quantify how our results change if the gas distribution surrounding the DCBH were flattened (or -- say -- in a thick disk). This affects the predicted Ly$\alpha$ {\it luminosity} in several ways:
\begin{itemize}[leftmargin=*]
\item Ionizing radiation can escape in certain directions, but not in others. If we denote the solid angle along which ionising photons escape with $\Omega_{\rm ion}$, then the total production rate of Ly$\alpha$ photons inside the cloud is reduced by a factor of $(1-\Omega_{\rm ion})$. 
\item Ly$\alpha$ recombination photons will likely scatter, and efficiently escape along the same paths as the ionising photons, thus leading to a slight beaming of Ly$\alpha$ flux \citep[see][]{Behrens14}. Similarly, Ly$\alpha$ photons produced inside the HI gas (following collisional excitation) escape in directions of low HI column densities. This beaming will partially compensate\footnote{Or possibly overcompensate. The amount of beaming depends on $\Omega_{\rm ion}$, the gas kinematics etc. Beaming can boost the flux in certain directions, but by a factor of $\sim$ a few at most \citep{Behrens14}.} the reduced Ly$\alpha$ production rate.
\item Anisotropic escape of Ly$\alpha$ along low column density directions will suppress the importance of collisional de-excitation at a fixed density, and will lead to collisional deexcitation becoming important at higher densities than what is shown in Figure~\ref{fig:lum}.
\end{itemize}

We thus expect our predicted maximum Ly$\alpha$ luminosity to be practically the same for more realistic gas distributions, but that they likely apply for a wider range of densities and a subset of viewing angles (\textcolor{black}{also see \S~\ref{sec:real}}). 
The assumed geometry has a bigger impact on the predicted Ly$\alpha$ {\it spectrum}. The biggest impact of geometry occurs when the ionized region `breaks' out of the cloud only in certain directions. We refer to sightlines to the DCBH that do not encounter any neutral gas as density-bound sight lines, while sightlines that do intersect neutral hydrogen as ionization bound sight lines.  Here we quantify how geometry affects the emerging spectra in more detail.

We place a fully ionized sphere with radius $R_{\rm ion}$ within a ellipsoid of neutral gas on a Cartesian grid of $256^3$ cells in a box with box-size of $(100,\,100,\,20)\,$pc. The ellipsoid is characterized by three half-axis given by $R_{{\rm ell}, a}=R_{{\rm ell},b} = 50\,$pc and $R_{{\rm ell},c}=10\,$pc. We focussed on the two following setups: \textit{(i)} the ionized sphere is touching the surface of the ellipsoid ($R_{\rm ion} = R_{{\rm ell}, c}$). We refer to this model as the `no break-out' model \textit{(ii)} the ionized sphere breaks out of the ellipsoid (i.e. $R_{\rm ion} = 2 R_{{\rm ell.}, c}$), which we call `break-out' model. We chose the hydrogen number density so that the maximum column density between the center and the outside is $10^{22}\,{\rm cm}^{-2}$, i.e., $n_{\rm HI}(R_{{\rm ell},a} - R_{\rm ion}) = 10^{22}\,{\rm cm}^{-2}$. We assume that the gas temperature is $T=10^4$ K.

In both models, we place three different sources: a central source (exponential distribution with scale length $0.1\,$pc, this emission gives rise to the `central spectrum'), a uniform distribution inside the ionized region (the `inner spectrum'), and a uniform distribution outside the ionized region (the `outer spectrum'). The intrinsic spectrum is a Gaussian with $\sigma = 12.9$\kms in all cases. The inner/central and outer spectra can be interpreted as being associated with Ly$\alpha$ emission powered by recombination and X-ray/gravitational heating, respectively. Fig.~\ref{fig:ellipsoid} shows the resulting spectra as seen by an observer viewing the ellipsoid face-on or edge-on, prior to processing of these spectra by intergalactic radiative transfer. The figure also includes sketches of the geometrical setups in which the emitting regions are marked in the same color as the corresponding spectrum. Fig.~\ref{fig:ellipsoidigm} shows how intergalactic radiative transfer alters these spectra. Our main results are

\begin{itemize}[leftmargin=*]
\item The spectra are doubled peaked and broad in the no break-out scenario. The width of the spectra depends only weakly on viewing angle and emission site. The width of the spectrum is set approximately by the lowest column density through the flattened cloud. One way to see this is that the HI column density along the shortest axis ($R_{\rm ell},a$) is $\sim \frac{R_{\rm ell},a}{R_{\rm ell}.c}\times10^{22}$ cm$^{-2}\sim 2\times 10^{21}$ cm$^{-2}$. This column density translates to a line center optical depth $\tau_0 \approx 1.2\times 10^8(T/10^4\hs{\rm K})^{-0.5}$. For a spherical cloud with this line center optical depth, we would expect Ly$\alpha$ spectra to exhibit peaks at $\Delta v \pm v_{\rm th}(a_v\tau_0)^{1/3}\approx \pm 490 (T/10^4\hs{\rm K})^{1/6}$ km s$^{-1}$ (where $a_v=4.7\times 10^{-4}(T/10^4\hs{\rm K})^{-0.5}$ denotes the Voigt parameter), which is close to the true location of the peaks. Ly$\alpha$ photons thus diffuse outward in real-space until they reach the edge of the cloud. The typical column density of HI gas they encountered corresponds to the column density associated with the shortest distance to the edge of the cloud, i.e. the photons typically escape along the paths of `least resistance'. The emerging spectra depends only weakly on viewing angle simply because the escape direction of photons is set by the last-scattering event, which does not depend on cloud geometry. The spectral shape depends weakly on where the Ly$\alpha$ photons were emitted.

\item The previous discussion, which highlighted that Ly$\alpha$ photons follow paths of least resistance, also allows us to understand results from the break-out models. The presence of low column density ($N_{\rm HI}< 10^{17}$ cm$^{-2}$) sightlines provides escape routes for Ly$\alpha$ photons, which do not require (and/or allow) them to diffuse far into the wings of the line profile. Indeed, the spectra associated with break-out models are typically narrow (peak separation $\sim 100$\,km\,s$^{-1}$). In addition to the characteristic double peaks, some spectra exhibit a third peak around line-center. The importance of this component depends strongly on viewing angle. In particular, the single peaked component is due to photons encountering no (or very little) neutral hydrogen along their path and is thus only present if a direct sight-line to the source exists and weakened with $|\mu|\rightarrow 0$.

\end{itemize}

\subsection{Ellipsoidal Clouds, with IGM}
\label{sec:geom_orient_IGM}

Fig.~\ref{fig:ellipsoidigm} shows how the IGM eliminates the blue peaks from our predicted spectra, which causes the spectra to appear redshifted. As we noted earlier, it is therefore difficult to distinguish clouds collapsing onto a DCBH from ordinary star forming galaxies based on the Ly$\alpha$ spectral line alone, especially when the spectrally broad component is subdominant to the narrow component. However, the presence of broad redshifted wings (extending to $\gsim 1000$ km s$^{-1}$ redward of line center), and possibly a second red-shifted peak, may be indicative of diffusion of Ly$\alpha$ photons through extremely opaque, dust-free, atomic hydrogen gas which is associated with DCBH formation. Interestingly, the fact that intergalactic radiative transfer only weakly affects photons that emerge from the cloud far in the red wing of the line profile (also see Dijkstra \& Wyithe 2010) may make it easier to detect these objects deep into the Epoch of Reionization.

\subsection{More Complex Gas Distributions}\label{sec:real}
\textcolor{black}{Our previous analyses and discussion applies to scattering through smooth, metal-free gas, which represent key properties that enabled DCBH formation. The situation becomes more complicated once the halo that hosts the DCBH merges with other (likely polluted) halos, and/or when feedback from accretion onto the black hole affects the gas hydrodynamically, and/or when in situ star formation starts to occur. When these more complicated processes occur, we are no longer in the regime where Ly$\alpha$ transfer can be addressed from first principles, and our predictions naturally no longer apply. }

\textcolor{black}{Anticipated changes to our predicted spectra include: ({\it i}) the presence of metals, and therefore, dust reduces the Ly$\alpha$ escape fraction from the clouds; ({\it ii}) the presence of complex gas flows, gas fragmentation and clumping generally increases Ly$\alpha$ escape compared to homogenous gas distributions that we adopted in our models. These processes would likely boost $f_{\rm esc}(n,T)$ at a given mean density. This may increase the Ly$\alpha$ cooling luminosity as Ly$\alpha$ photons may escape from denser gas. The anticipated boost is difficult to predict as fragmentation will lead to in-situ star formation, which in turn leads to stellar feedback occurs. At this point, ab initio predictions for the emerging Ly$\alpha$ spectra become unreliable.}

 \section{Comparison to CR7}
\label{sec:shell_model}

\subsection{Luminosity}
CR7 is a bright Ly$\alpha$ emitting source at $z\sim 6.6$ which has been associated with a DCBH \citep{Sobral15,Pallottini15,AgCR7}. The total Ly$\alpha$ flux observed from CR7 implies a Ly$\alpha$ luminosity of $L_{\alpha}=10^{44}$ erg s$^{-1}$. The total {\it intrinsic} Ly$\alpha$ luminosity is likely higher, as the observed Ly$\alpha$ luminosity has been suppressed by intergalactic scattering, and possibly by interstellar dust. \citet{AgCR7} argue that in order to reproduce the observed ratio of flux in the Ly$\alpha$ and He1640 lines, the observed Ly$\alpha$ luminosity only represents one sixth of the total intrinsic luminosity. Our calculations in \S~\ref{sec:lum} imply that in order to explain its observed Ly$\alpha$ luminosity, CR7 must be powered by a black hole with mass $M\gsim 10^7 M_{\odot}$.

\textcolor{black}{Previous works have attributed the Ly$\alpha$ luminosity of CR7 to black holes with mass $M_{\rm BH}\sim$ a few $\times 10^6 M_{\odot}$ \citep[e.g.][]{Pallottini15,AgCR7}. 
The {\it bolometric} Eddington luminosity of a  $M_{\rm BH}\sim 10^6 M_{\odot}$ black hole equals $L_{\rm bol}\sim 1.3 \times 10^{44}$ erg s$^{-1}$, which makes it problematic to account for the total observed Ly$\alpha$ luminosity of CR7 of $L_{\alpha} \sim 10^{44}$ erg s$^{-1}$ with (sub)Eddington accretion. This problem worsens when the intrinsic Ly$\alpha$ flux is required to be $\sim 6$ times larger (see above)\footnote{\textcolor{black}{Boosting $F_{{\rm Ly}\alpha}$ (see Eq~\ref{eq:reclya}) further enhances the Ly$\alpha$ luminosity at fixed $M_{\rm BH}$. Physically this corresponds to requiring that high-energy photons enhance the Ly$\alpha$ production rate. However, the Ly$\alpha$ luminosity remains bound by $L_{\alpha}\leq f_{\rm heat}f_{\rm X}L_{\rm edd}$. To reach $L_{\alpha}\sim 10^{44}$ erg s$^{-1}$ with $M_{\rm BH} < 10^7 M_{\odot}$ requires $f_{\rm heat}f_{\rm X}>0.1$, or even $f_{\rm heat}f_{\rm X}\gsim 0.6$ when constraints for the ratio of He1640 to Ly$\alpha$ line fluxes are considered. Especially this last constraint is unphysical.}}}.

\subsection{Ly$\alpha$ Spectrum}\label{sec:cr7spec}

\begin{figure*}
\centering
\includegraphics[width=14.0truecm,angle=0]{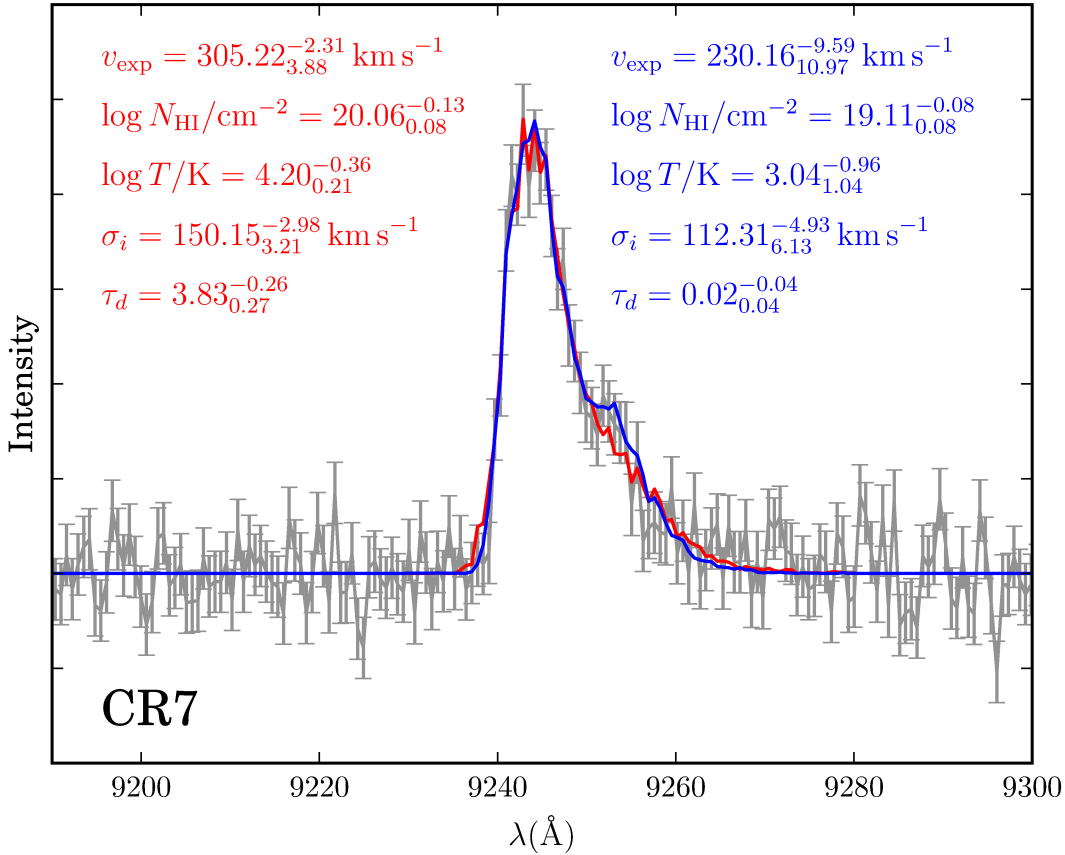}
\caption{Ly$\alpha$ spectrum of CR7 (Sobral et al. 2015). The {\it solid lines} represent our best-fit shell models. The {\it red solid line} shows our best-fit shell model that we obtain by only using a prior on CR7's redshift $z$. The {\it blue solid line} shows our best-fit model if we force the scattering medium to have a low dust content by adding a prior on the dust content (see text). The best-fit shell model parameters for both models are shown. Shell-model fitting shows that the observed Ly$\alpha$ spectrum of CR7 favors the Ly$\alpha$ source to be surrounded by a scattering medium with a column density of $\log [N_{\rm HI}/{\rm cm}^{-2}]\sim 19-20$ which is outflowing at a rate $v_{\rm exp} \sim 200-300$ km s$^{-1}$. These parameters do not differ drastically from those inferred for lower redshift Ly$\alpha$ emitting galaxies, and may suggest that the interstellar medium of CR7 is already shaped by star formation and stellar feedback.}
\label{fig:shell_model_fit}
\end{figure*}
\textcolor{black}{CR7s rest-frame ultra-violet spectrum was obtained with both X-SHOOTER on the VLT and DEIMOS on Keck (see Sobral et al. 2015) with slits of 0.9'' and 0.75'' respectively and resolutions of $R\sim7500-10000$, and with a total integration time of 3.75 hours. As detailed in Sobral et al. (2015), no continuum is detected either blue-ward or red-ward of Ly$\alpha$ (but rest-frame Lyman-Werner continuum is detected at rest-frame 916-1017\,\AA; see Sobral et al. 2015).}

The observed FWHM of the Ly$\alpha$ line from CR7 is FWHM$\sim 260$ km s$^{-1}$ (Sobral et al. 2015), which is a factor of $\sim 2$ smaller than in the models with $N_{\rm HI}=10^{22}$ cm$^{-2}$ (see \S~\ref{sec:spec}). Analytic considerations (Eq~\ref{eq:FWHM}) imply that this FWHM thus translates to an HI column density in the scattering medium that is $\sim 2^3$ times smaller (i.e. $\log [N_{\rm HI}/{\rm cm}^{-2}]\sim 21$) to explain the observed FWHM. We show below that the actual column density of the scattering medium is likely well below this. The required HI column density, $\log [N_{\rm HI}/{\rm cm}^{-2}]\ll 22$, in these models implies a very specific evolutionary state in which the cloud is mostly ionized, and surrounded by a thin skin of HI gas.
 
In previous sections we predicted Ly$\alpha$ spectra associated with the DCBH formation scenario. We compared these predictions to observations to CR7 in several places. Here, we perform a more quantitative analysis of the observed spectrum of CR7 (Sobral et al. 2015) by fitting `shell models' to the data. The shell model represents another class of spherically symmetric models that has been routinely adopted to describe Ly$\alpha$ scattering on interstellar scales. The shell model is a simple, six-parameter model which has been proven to be surprisingly successful in reproducing observed Ly$\alpha$ spectra \citep[see, e.g.,][]{Ahn2004ApJ...601L..25A,Verhamme2008}. The model consists of a central Ly$\alpha$ and continuum emitting source surrounded by a shell of outflowing hydrogen (and dust). The source can be characterized by the Ly$\alpha$ equivalent width $EW_i$ and the width of the intrinsic line $\sigma_i$. The shell content is described fully by its hydrogen column density $N_{\rm HI}$, the dust optical depth $\tau_d$, the (effective) temperature $T$ (or, equivalently by the Doppler parameter $b$), and, the outflow velocity $v_{\rm exp}$. Here, we constrain the shell-model parameters that provide the best description of the spectrum of CR7.
 
 We use the pipeline described in Gronke et al. (2015): a set of $10,800$ pre-computed shell model spectra in combination with a post-processing procedure allows us to obtain the best-fit values, and also the uncertainties \& potential degeneracies between the parameters. \textcolor{black}{The results of the fitting procedure are shown in Figure~\ref{fig:shell_model_fit} as one- and two-dimensional projections of the posterior likelihood distribution sampled by a Monte-Carlo sample. We use the affine-invariant Monte-Carlo sampler \texttt{emcee} \citep{Foreman-Mackey2012} with $1600$ steps and $400$ walkers. We do not fit for the Ly$\alpha$ equivalent width $EW_i$ since no continuum was detected \citep{Sobral15}. Instead, we simultaneously fit for the redshift $z$ using a Gaussian prior with $(\mu,\sigma)=(6.604,0.003)$}.

\textcolor{black}{The {\it red solid line} in Figure \ref{fig:shell_model_fit} shows the best-fit model to the data. The best-fit shell model parameters and their uncertainties are listed in the top-left corner. Two parameters are of particular interest: we obtain $\log_{10}N_{\rm HI}/{\rm cm}^{-2}=20.06^{+0.08}_{-0.13}$ and $\tau_d=3.83^{+0.27}_{-0.26}$. These parameters differ from the values we would naturally associate with DCBH formation (namely, $\tau_d = 0$ and $N_{\rm HI} \gg 10^{20}\,{\rm cm}^{-2}$).} This tension is qualitatively\footnote{\textcolor{black}{Analytic considerations prefer $\log [N_{\rm HI}/{\rm cm}^{-2}]\sim 21$, which is because these assume that the Ly$\alpha$ photons are all initially emitted at line center. In contrast, in the best-fit shell-models the initial Ly$\alpha$ spectrum is a Gaussian with $\sigma\approx 150$ km s$^{-1}$. The spectral distribution of Ly$\alpha$ line is therefore broad to begin with, which then requires a smaller HI column density to broaden the line to match the observations. The fact that formally the data prefers a high-$\sigma$, low $N_{\rm HI}$ solution is entirely due to the spectral shape of the line.}} consistent with our analysis of the spectra emerging from contracting gas cloud, in which the observed width of the Ly$\alpha$ line favored low HI column densities. The high value of $\tau_d$ is preferred due to the sharp drop-off of the spectrum on the blue side. \textcolor{black}{However, this high value is inconsistent with the upper limit on the metallicity of the gas of $Z \lsim 0.02 Z_{\odot}$ found by Hartwig et al. (2016) due the non-detection of the CIII] line, and, results in a very low escape fraction of $f_{\rm esc}\lsim 10\%$. Due to these tensions we have repeated our fitting procedure adding a narrow prior on $\tau_d$ ($(\mu,\sigma)=(0,0.05)$) which yields $\tau_d=0.02^{+0.04}_{-0.04}$. In this case, the line profile favors an even lower column density\footnote{We did not include the IGM in our fitting procedure. We have also repeated our fitting procedure only for the data points redward of $9246$\,\AA (i.e. ignoring the wavelengths which are likely affected by IGM absorption). This fitting procedure -without putting a prior on the dust content - favors even lower HI column densities of $\log N_{\rm HI}/{\rm cm}^{-2}\sim 18-20$.} of $\log N_{\rm HI}/{\rm cm}^{-2}= 19.11^{+0.08}_{-0.08}$.}

The inferred shell-model parameters lie within the range of values inferred for lower redshift Ly$\alpha$ emitters: preferred shell column densities in $z\sim 2.2$ Ly$\alpha$ emitters are $\log [N_{\rm HI}/{\rm cm}^{-2}]\sim 19$ (Hashimoto et al. 2015), and in green pea galaxies at $z\sim 0.3$ are $\log [N_{\rm HI}/{\rm cm}^{-2}]\sim 19-20$ (Yang et al. 2015). The required shell velocity is a higher for CR7 than in {\it most} lower redshift galaxies  \textcolor{black}{(but see Vanzella et al. 2010 for an example of a galaxy at $z\sim 5.6$ where the shell velocity and HI column density are both higher). The scattering medium of CR7 therefore does not seem drastically different than in lower redshift Ly$\alpha$ emitting galaxies.} 

\textcolor{black}{This implies that if CR7 hosts a DCBH, that then the conditions that enabled DCBH formation may have been mostly erased. This may not be surprising as the required black hole mass of $M\gsim 10^7 M_{\odot}$ exceeds the mass of DCBHs when they first form \citep[which is $\sim 10^5-10^6 M_{\odot}$ e.g.][]{Latif13,Shlosman16}. This requires that the black hole has grown significantly since it first formed (see Hartwig et al. 2016, Agarwal et al. 2016), which in turn implies that the host galaxy likely merged since the BH formed.
It is then difficult to {\it prove} that CR7 is indeed powered by a BH that formed via direct collapse, or instead via a BH that grew from a stellar mass seed. Instead, Hartwig et al. (2016) show that the non-detection of the CIII] line puts an upper limit on the metallicity of the gas at $Z \lsim 0.02 Z_{\odot}$, and that this condition on metallicity makes it more {\it likely} for the BH to have first formed as a massive seed via direct collapse.}

\subsection{Constraints from the He1640 Line}
\textcolor{black}{As we mentioned earlier, the equivalent width of the He1640 line provides a direct measure of the hardness of the source illuminating the gas cloud\footnote{\textcolor{black}{It is interesting to point out that the detection of He1640 line formally directly rules out models that purely invokes gravitational heating, though these models were also ruled out based on their predicted Ly$\alpha$ luminosity.}} (e.g. Sobral et al. 2015, also see Johnson et al. 2011). In addition, constraints imposed by the ratio of line fluxes in He1640 to Ly$\alpha$ have been explored in previous works (Johnson et al. 2011, Sobral et al. 2015, Agarwal et al. 2016, Hartwig et al. 2016). The He1640 line provides a third constraint, as it constrains both the systemic velocity of CR7 and the width of non-resonant nebular lines, and thus the width of the Ly$\alpha$ spectral line prior to scattering. This information  in turn constrains the Ly$\alpha$ transfer process.} 

\textcolor{black}{It is interesting that our best-fit model favor the intrinsic FWHM of the Ly$\alpha$ line to be $\sim 2.35\times \sigma_i\sim 250$ km s$^{-1}$, which is a factor of $\sim 2$ broader than the FWHM of the He1640 line. This problem is also encountered for Ly$\alpha$ emitters at $z\sim 2.2$ which also required $\sigma_i$ to be larger than the observed width of the H$\alpha$ line (see Hashimoto et al. 2015), and is still not understood.}

\section{Discussion \& Conclusions}
\label{sec:conc}

We have modelled the relevant radiative processes of the Ly$\alpha$ line through simplified representations of the `direct collapse black hole' (DCBH) scenario. 
The suppressed gas fragmentation, the absence of star formation and stellar feedback - all key characteristics of the DCBH formation scenario - simplify the Ly$\alpha$ radiative transfer problem, which is known to depend sensitively on all these processes. Our main results are: 

{\bf 1.} Gravitational heating of the collapsing cloud gives rise to a Ly$\alpha$ cooling luminosity of up to $\sim 10^{38}(M_{\rm gas}/10^6\hs M_{\odot})^2$ erg s$^{-1}$. Though gravitational heating can increase the Ly$\alpha$ production much further during the later stages of collapse, collisional deexcitation efficiently suppresses the emerging Ly$\alpha$ flux. During these later stages, the cloud may be visible through its two-photon continuum emission \citep{Dijkstracool,Inayoshi15b}, or through stimulated 3-cm emission (Dijkstra et al. 2016). 

{\bf 2.} Photoionization of the cloud by a central source can boost the clouds Ly$\alpha$ luminosity to values as large as $L_{\alpha} \sim 10^{43}(M_{\rm BH}/10^6 M_{\odot})$ erg s$^{-1}$ (irrespective of $M_{\rm gas}$) over a limited range of densities. Here, $M_{\rm BH}$ denotes the mass of the black hole powering this source. For lower densities the ionized gas is density-bound, which reduces the conversion efficiency of ionising radiation into Ly$\alpha$. For higher densities the cloud is ionization-bound, and the collisional de-excitation in the neutral gas again reduces the Ly$\alpha$ luminosity. \textcolor{black}{This luminosity, and that for gravitational cooling, are not affected by our adopted simplified geometry, and are set entirely by the energetic considerations.}

{\bf 3.} We computed Ly$\alpha$ spectra for a suite of spherical and flattened collapsing gas clouds. The predicted spectra are double peaked, with an enhanced blue peak. The width and velocity off-sets of the peaks range from a few tens to few thousands km s$^{-1}$, and are set by the HI column density through the cloud in close agreement with analytic predictions. We process the emerging spectra through the intergalactic medium (IGM), and find that the IGM completely eliminates the blue peak. The resulting spectra are asymmetric, and redshifted, and can look remarkably like those typically associated with scattering through a galactic outflow. However, the predicted velocity off-set and FWHM can exceed $\sim 1000$ km s$^{-1}$, which is larger than what is typically found for high-redshift Ly$\alpha$ emitting galaxies \textcolor{black}{, and thus possibly an interesting observational signature}.

{\bf 4.} In flattened clouds the predicted FWHM and overall off-set of the Ly$\alpha$ spectral line is set closely by the minimum HI column density associated with the cloud. An example of this consists of the extreme case in which a central source ionized the cloud in certain directions, but not others.  In this case, the predicted Ly$\alpha$ spectral line is narrow, with flux emerging at line center. Moreover, the emerging spectrum and the fraction of the flux that emerges at line center depends on viewing angle. Specifically, sightlines that allow ionizing photons to escape are associated with most Ly$\alpha$ photons escaping at line center (also see Behrens et al. 2014, Verhamme et al. 2015).

{\bf 5.} We compared to observations of CR7, a luminous Ly$\alpha$ emitter at $z\sim 7$ which is potentially associated with a DCBH (Sobral et al. 2015). Our analysis implies that the Ly$\alpha$ line luminosity of CR7 alone implies that if it is indeed associated with a DCBH, then it must be associated with a black hole of mass $M_{\rm BH}\gsim 10^7 M_{\odot}$. If CR7 is indeed powered by such a massive black hole, then it must have accreted most of this mass after its formation, as the maximum mass of a newly formed DCBH is $\sim 10^5-10^6 M_{\odot}$ \citep[e.g.][]{Latif13,Shlosman16}. Supply of this gas to the black hole required mergers with other nearby halos \citep{AgCR7,Hartwig16}. These mergers would likely have erased the very specific conditions in the gas that were required for DCBH formation. This may explain that the spectral line shape of CR7 favors a low HI-column density, $\log [N_{\rm HI}/{\rm cm}^{-2}]\approx 19-20$, outflowing scattering medium. These parameters are consistent with those inferred for lower redshift Ly$\alpha$ emitting galaxies, which may reflect that the interstellar medium of CR7 is already shaped by star formation and stellar feedback. 

Alternatively, our radiative transfer calculations through simplified representations of the DCBH formation process can reproduce the observed the spectral line shape of CR7 when the gas cloud is in a very specific evolutionary state (again associated with a low column density of HI), and that the intergalactic medium eliminated the blue side - which contains most flux - of the emitted Ly$\alpha$ flux out of our line-of-sight. In this case, CR7 should be surrounded by an Ly$\alpha$ halo \textcolor{black}{that extends much further than the present observations} with a luminosity (significantly) larger than $10^{44}$ erg s$^{-1}$.

\textcolor{black}{The implication that the interstellar medium of CR7 may already be shaped by star formation and stellar feedback begs the question whether these stars can be a `pure' $\sim 3$ Myr old Population III stellar population (as discussed by Sobral et al. 2015, Visbal et al. 2016). It is then interesting to consider that the Ly$\alpha$ spectral line shape favors scattering through outflowing gas (see \S~\ref{sec:cr7spec}), which implies that its ISM is being enriched, and that enough time must have elapsed for the effects of feedback to have commenced. If the Ly$\alpha$ line was actually intrinsically symmetric and the observed asymmetry is entirely due to the IGM (as discussed above), then we again expect CR7 to be surrounded by an extended, still undetected Ly$\alpha$ halo with a luminosity (significantly) larger than $10^{44}$ erg s$^{-1}$.}

Finally, the association of broad redshifted Ly$\alpha$ lines with the DCBH formation scenario makes this Ly$\alpha$ flux less susceptible to scattering in the IGM compared to narrower emission lines that have been observed from star forming galaxies \citep{DW10}. This reduced sensitivity to the neutral IGM may increase the likelihood of finding DCBHs among the high-redshift Ly$\alpha$ emitting population: a source with an apparent Ly$\alpha$ luminosity of $\sim 10^{42}$ erg s$^{-1}$ at $z=10$ (corresponding to a flux of $0.8 \times 10^{-18}$ erg s$^{-1}$ cm$^{-2}$) can be detected with the James Webb Space Telescope (JWST) in $10^4$ s at $\sim 3-\sigma$ significance\footnote{\url{http://www.stsci.edu/jwst/science/sensitivity/spec1.jpg}}. This - combined with the increased sensitivity of NIRSPEC towards longer wavelengths - suggests that JWST at least has the sensitivity to detect broad Ly$\alpha$ lines from $M_{\rm BH} \gsim 10^5 M_{\odot}$ DCBHs out to redshift well beyond $z=10$.

% ***************************************************************************
\section*{Acknowledgments} 
MD thanks the Institute of Astronomy at the University of Edinburgh and the astronomy department at UCSB for their kind hospitality. MG thanks the Physics \& Astronomy department at JHU for their hospitality. DS acknowledges financial support from the Netherlands Organisation for Scientific research (NWO) through a Veni fellowship. We thank an anonymous referee for a critical, helpful report that improved the content of this paper. Based on data products from observations made with ESO Telescopes at the La Silla Paranal Observatory under ESO programme ID 294.A-5018Ó  (X-SHOOTER data).

% **************************************************************************
% BIBLIOGRAPHY

%\bibliographystyle{./style/apj}
%\bibliography{./style/ref}
\appendix
\section{More Details on the Calculations}
\subsection{Temperature Dependence Ionized Fraction}
\label{app:t}

Collisional ionization equilibrium forces the recombination rate to equal the collisional ionization rate, i.e. $n_en_p\alpha_{\rm B}(T)=n_e n_{\rm HI} q_{\rm ion}(T)$. Here, $\alpha_{\rm B}(T)$ is the case-B recombination coefficient and $q_{\rm ion}$ denotes the collisional ionization rate coefficient. The equilibrium value is
\begin{equation}
x_e\equiv \frac{n_e}{n}=\frac{n_p}{n}=\frac{q_{\rm ion}(T)}{\alpha_{\rm B}(T)},
\end{equation} where we have taken fitting formulae for $\alpha_{\rm B}(T)$ and $q_{\rm ion}(T)$ from \citet{Hui97}.

\subsection{Temperature Evolution of Cloud}
\label{app:intermediate}

\begin{figure*}

\includegraphics[width=6.0cm,angle=270]{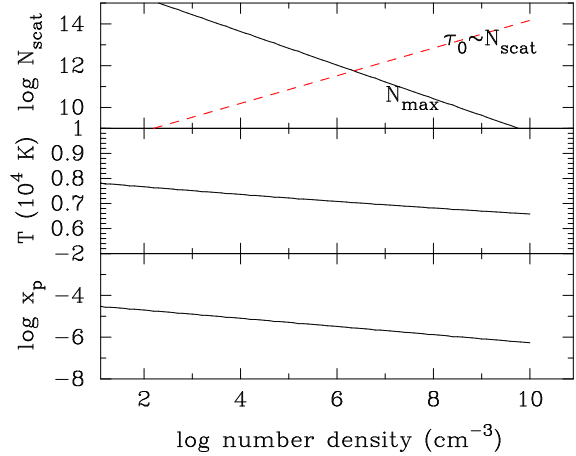}
\centering
\caption{This plot shows density dependence of various quantities associated with a collapsing gas cloud of uniform density that is heated by the gravitational collapse. The {\it top panel} shows the line centre optical depth $\tau_0$ - which equals (approximately) the average number of scattering events a Ly$\alpha$ photon undergoes before escaping - as a function of $n$ ({\it black solid line}). The {\it red dashed line} shows the maximum allowed number of scattering events, before collisional deexcitation becomes important. The two are equal at $n\sim 10^6$ cm$^{-3}$, which implies that at higher densities, collisional deexcitation suppresses the emerging Ly$\alpha$ flux. The {\it central/bottom panel} shows that the temperature/residual ionised fraction decreases towards higher density.} 
\label{fig:uniformcloud}
\end{figure*}

In \S~\ref{sec:lum} we computed the Ly$\alpha$ luminosity, and the Ly$\alpha$ escape fraction as a function of density $n$. Figure~\ref{fig:uniformcloud} show some intermediate results of these calculations. The {\it red dashed line} in the {\it upper panel} shows the line centre optical depth $\tau_0=N_{\rm HI}\sigma_0$ through the cloud, where $\sigma_0=5.88\times 10^{-14}(T/10^4\hs{\rm K})^{-1/2}$ cm$^2$ denotes the absorption cross-section at line centre. The column density increases as $N_{\rm HI} \propto n^{2/3}$, and therefore $\tau_0$ also increase {\it almost} as $\tau_0 \propto n^{2/3}$. The reason that $\tau_0$ does not increase exactly as $\tau_0 \propto n^{2/3}$ is because of the mild temperature dependence of $\sigma_0$ (as discussed below and shown in the {\it central panel} of Fig~\ref{fig:uniformcloud}, the temperature changes with $n$). Finally, the maximum number of scatterings $N^{\rm max}_{\rm scat} \propto n^{-1}_{\rm p}$, and therefore decreases with overall number density $n$ as the ionised fraction changes quite slowly with density $n$ (see the {\it lower panel}). The total number of scatterings equals the maximum at $n\sim 10^5$ cm$^{-3}$. This implies that at higher densities, collisional deexcitation suppresses the emerging Ly$\alpha$ flux, which was also apparent in Figure~\ref{fig:lum}. 

The {\it central panel} shows the temperature evolution of the cloud with $n$, under the assumption that the gravitational heating rate is balanced by cooling via collisional excitation ($\sim 40\%$ of which is Ly$\alpha$ cooling). The gravitational heating rate increases towards higher densities as $\propto n^{2/3}$ (see Eq~\ref{eq:gravheat}), but the collisional cooling rate increases slightly faster (as $n^2$), which is compensated by a reduction in temperature towards higher $n$ (because the collisional cooling rate is so sensitive to the gas temperature, this decrease in $T$ is only small). Finally, the {\it lower panel} shows how the residual ionised fraction also decreases with $n$ as a result of the decreasing temperature $T$. Specifically, $x_{\rm p}$ changes by $\sim 1$ order of magnitude over $\sim 8$ orders of magnitude in $n$. This implies that $x_{\rm p} \propto n^{-0.12}$, and that therefore $n_{\rm p}=nx_{\rm p} \propto n^{y}$, where $y=0.88$. We used this dependence in the paper.

\subsection{Isothermal Density Profiles}
\label{app:iso}
Throughout the paper we assumed isothermal density profile. This greatly simplified our analysis, which improved the clarity of the presentation and which allowed us to apply analytic results obtained for Ly$\alpha$ transfer. Here, we repeat some of our calculations for a cored isothermal density profile, which is a more realistic description of the density field (at least in 1D hydrodynamical simulations, see Pacucci \& Ferrara 2015). This density profile is
\begin{eqnarray}
\rho(r)= \left\{ \begin{array}{ll}
         \frac{A}{1+(r/r_c)^2} & {\rm for}\hs\hs\hs r \leq R;\\
         0 & {\rm for}\hs\hs\hs r > R\end{array} 
         \right.
\end{eqnarray} where $A$ is a normalisation constant, and $r_c$ is the core radius which we assume to be $r_c=0.1R$, where $R$ denotes the edge of the cloud (as was the case for the uniform cloud). The total mass enclosed within radius $r$ is
\begin{equation}
M_{\rm gas}(<r)=\int_0^{r}dr\hs 4\pi r^2 \frac{A}{1+(r/r_{\rm c})^2},\hs\hs\hs A=\frac{M_{\rm gas}}{4\pi r_{\rm c}^3\left( \frac{R}{r_{\rm c}}-\arctan\frac{R}{r_{\rm c}}\right)}.
\end{equation} 

We can express the HI column density through the cloud as
\begin{eqnarray}
N_{\rm HI}=\frac{Ar_{\rm c}}{\mu m_{\rm p}} \arctan\left (\frac{R}{r_{\rm c}}\right)=\frac{100R\hs \mathcal{K}(R/r_{\rm c})}{3}\bar{n}\approx 5.5\bar{n}R, &  \hspace{5mm}\mathcal{K}(x)=\frac{\arctan x}{x-\arctan x},
\end{eqnarray} where we plugged in $R/r_{\rm c}=x=10$, which translates to $\mathcal{K}(x) \sim 0.17$. That is, for a fixed cloud size and mass, the HI column density through a cored isothermal profile (with $r_{\rm c}=0.1R$) is about 5.5 larger than in the uniform case.

The total recombination rate in turn can be expressed as
\begin{eqnarray}
\dot{N}_{\rm rec}^{\rm iso}=\alpha_{\rm B}\frac{M^2_{\rm gas}}{V(\mu m_{\rm p})^2}\mathcal{N}(R/r_c)=\dot{N}_{\rm rec}^{\rm uni}\mathcal{N}(R/r_c)\approx 6.7\dot{N}_{\rm rec}^{\rm uni}, &  \hspace{5mm}\mathcal{N}(x)=\frac{x^3}{3}\frac{\arctan x -\frac{x^2}{x^2+1}}{(x-\arctan x)^2},
\end{eqnarray} where again we plugged in $x=10$. The total recombination rate for a given gas mass \& size is thus $\sim 7$ times larger than in the uniform case. We stress that this does not boost the maximum Ly$\alpha$ luminosity from the cloud, as this is set by the total production rate of ionising photons. The enhanced recombination rate at given average density would only cause the total recombination luminosity to saturate at a lower density than what was shown in Figure~\ref{fig:lum}.

The gravitational binding energy can be written as
\begin{eqnarray}
U_{\rm bind}=-\frac{GM^2_{\rm gas}}{R}\mathcal{F}(R/r_{\rm c}) \approx -0.8\frac{GM^2_{\rm gas}}{R}, & \hspace{5mm}
\mathcal{F}(x)=x\frac{\int_0^{x}du\hs\left(\frac{u^2}{1+u^2} -\frac{u \arctan u}{1+u^2}\right)}{\left(x-\arctan x\right)^2}.
\end{eqnarray} The total gravitational binding energy at fixed $M$ and $R$ is slightly more negative than in the uniform case, which is because the gas has settled deeper in the gravitational potential well compared to the uniform density case.

\newpage 
\label{lastpage} 
\end{document}